\documentclass[aps,preprintnumbers,nofootinbib]{revtex4}
\usepackage{amsmath,amssymb,mathtools}
\usepackage{graphicx}
\usepackage{hyperref}
\usepackage{subfigure}
\usepackage{comment}
\usepackage{tensor}
\usepackage{cases}
\usepackage{color}
\usepackage{soul}
\usepackage{bm}
\usepackage{filecontents} 

\newcommand{\red}[1]{{\textcolor{black}{#1}}}

\newcommand{\be}{\begin{eqnarray}}
\newcommand{\ee}{\end{eqnarray}}

\newcommand{\simgt}{\lower.5ex\hbox{$\; \buildrel > \over \sim \;$}}
\newcommand{\simlt}{\lower.5ex\hbox{$\; \buildrel < \over \sim \;$}}

\begin{document}

\preprint{HUPD-1710}
\vspace{2cm}
\title{Gravitational Redshifts of Clusters and Voids}
\vspace{1cm}

\author{Daiki Sakuma}
\affiliation{Graduate School of Science, Department of Physics, 
Hiroshima University, Higashi-Hiroshima, Kagamiyama 1-3-1, 739-8526, Japan}

\author{Ayumu Terukina}
\affiliation{Graduate School of Science, Department of Physics,
Hiroshima University, Higashi-Hiroshima, Kagamiyama 1-3-1, 739-8526, Japan}

\author{Kazuhiro Yamamoto}
\affiliation{Graduate School of Science, Department of Physics, 
Hiroshima University, Higashi-Hiroshima, Kagamiyama 1-3-1, 739-8526, Japan}

\author{Chiaki Hikage}
\affiliation{Kavli Institute for the Physics and Mathematics of the Universe (Kavli IPMU, WPI), The University of Tokyo, 5-1-5 Kashiwanoha, Kashiwa, Chiba, 277-8583, Japan}

\begin{abstract} 
  We investigate gravitational redshifts (signals of gravitational potential)
  in measurements of the redshifts of cosmological objects, i.e., central and satellite
  galaxies in clusters of galaxies, intracluster gas, as well as galaxies associated with
  voids by developing simple theoretical models.
  In the analysis with satellite galaxies in clusters, we develop a very simple analytic
  model for satellite galaxies
  virialised in halos, which enables us to evaluate the
  signals depending on the properties of the halo occupation distribution of galaxies. 
  We obtain results consistent with recent previous results, though our results are
  restricted to the satellite galaxies inside the virial radius. 
  In the analysis of intracluster gas, we develop a simple analytic model including
  the effect of random motions of gases, which are assumed to generate nonthermal pressure.
  We demonstrate a possible contribution of the random motions of gases to gravitational
  potential measurements.
  We also investigate a possible signature of the gravitational potential in measurements
  of galaxies
  associated with voids by utilizing a simple analytic model.
  We show that the second-order Hubble term, which appears in the expansion of the scale
  factor around the centre of a void, may make a significant contribution depending on
  the way the galaxy samples are analysed.
  \red{The studies on the possible signals of gravitational potential for intracluster gases and voids
  are performed for the first time.}
\end{abstract} 

\maketitle

\def\rpara{{r_{\scriptscriptstyle \|}}}
\def\rperp{{r_{\scriptscriptstyle \bot}}}
\def\kpara{{k_{\scriptscriptstyle \|}}}
\def\kperp{{k_{\scriptscriptstyle \bot}}}
\def\spara{{s_{\scriptscriptstyle \|}}}
\def\sperp{{s_{\scriptscriptstyle \bot}}}
\def\qpara{{q_{\scriptscriptstyle \|}}}
\def\qperp{{q_{\scriptscriptstyle \bot}}}
\def\bfx{{\bm x}}

\begin{widetext}

\section{Introduction} 
\label{sec:intro}

Relativistic effects in the cosmological large-scale structure have been investigated 
by many authors
\cite{Baccanelli2016a,Bartolo2016,Baccanelli2016b,Bertacca2017,Yoo2014,Yoo2009,Alam1,Alam2}. 
Measurements of the gravitational redshifts\footnote{\red{We use the terminology,
    {\it gravitational redshift}, to express the redshift including the
    signal of gravitational potential dominantly,
    although such signals are contaminated by the second
    order Doppler velocities and other higher order effects of the general relativity. }}
 as a relativistic effect in measurements of
the redshifts of cosmological objects, have been 
reported recently using galaxies associated with clusters \cite{Wojtak,Zhao,Jimeno}.
Wojtak et al. first reported the detection of the gravitational redshift of satellite galaxies
in and around clusters \cite{Wojtak}. Zhao et al. pointed out the contribution of the transverse
Doppler velocity of galaxies to measurements of the gravitational potential \cite{Zhao}.
They also stressed the importance of the gravitational redshift as a test of general
relativity and modified gravity models.
Kaiser has pointed out that the nontrivial feature of the phase space distribution function of 
objects defined on the lightcone coordinate leads to an additional 
second-order Doppler term \cite{Kaiser}. 
Furthermore, other relevant effects in the measurements of the gravitational potential of
galaxies in clusters have been discussed \cite{Jimeno,Cai}.
Thus, gravitational redshift is a unique tool for testing the general relativity
and modified gravity theories. 

Motivated by these recent works, we investigate possible signatures of the gravitational 
redshift in clusters of galaxies and voids. 
We consider three systems. The first consists of satellite galaxies virialised in halos
of galaxy clusters, for which we derive a simple formula for the gravitational redshift
with the use of the halo occupation distribution (HOD) description with central galaxies and
satellite galaxies. \red{Our investigation is different from the previous works in the following point:}
Our analysis is restricted to scales within the virial radius of a halo, but our simple
analytic model is useful for understanding how the measurement of gravitational
redshift depends on the HOD properties of galaxy samples.
The second system concerns the gravitational redshift in observations of intracluster gas, which is
motivated by the recent precise measurement of intracluster gas motions in the Perseus
Cluster reported in Ref.~\cite{Hitomi}. A possible signature of the gravitational redshift
in measurements of intracluster gas is investigated. In this
analysis, we include the effect of random motions of the gas that explain the
nonthermal pressure predicted by numerical simulations. 
\red{ Possible contamination of random motions of gases to a measurement of gravitation potential is
pointed out for the first time.}
The third system comprises galaxies associated with voids. Because some galaxies
might be found inside voids, we may consider the possibility of measuring the gravitational
potential of voids. 
We investigate a possible signature of the gravitational potential of galaxies
associated with voids. \red{We stress that such an investigation is performed
for the first time as far as we know.}

This paper is organised as follows. 
In Sec.~II, we rederive a formula for the gravitational redshift given by Kaiser \cite{Kaiser},
starting from the geodesic equation. 
In Sec.~III, we investigate the gravitational redshift in galaxy samples in
redshift surveys, using the halo approach with the HOD with central galaxies and satellite galaxies. 
In Sec.~IV, we demonstrate a possible signature of the gravitational redshift in 
measurements of intracluster gas. 
In Sec.~V, we investigate a possible signature of the gravitational redshift
of galaxies associated with voids. We also show that the contribution from
a second-order Hubble term, which appears in the expansion of the scale
factor around the centre of a void, can be significant 
depending on the range of projecting galaxies in the line-of-sight direction
in the analysis. 
Section~VI is devoted to a summary and conclusions. 
In the Appendix, we summarise our theoretical modelling for the clusters of galaxies
in Sec.~IV. 

\section{Formulation} 
Detection of the gravitational redshift of galaxies in and near clusters of galaxies
has been reported in Refs.~\cite{Wojtak,Zhao,Jimeno}.
The theoretical formula developed by Kaiser is practical and useful \cite{Kaiser,Cai};
we rederive it, starting from the geodesic equation for a photon. 
We focus on the gravitational redshift in a system comprising a cluster of galaxies and 
voids, which are structures of much smaller than the horizon scale.
Therefore, the Newtonian gauge is useful and efficient; its line element is written as
\begin{eqnarray}
  &&ds^2=a(\eta)^2[-(1+2\psi)d\eta^2+(1+2\phi)d{\bm x}^2],
  \label{le}
\end{eqnarray}
where $a(\eta)$ is the scale factor as a function of the conformal time $\eta$, 
and $\psi$ and $\phi$ are the gravitational potential and the curvature potential, respectively. 
Up to the first order of $\psi$ and $\phi$, the geodesic equation for a photon 
leads to (see, e.g., \cite{Dodelson}) 
\begin{eqnarray}
{1\over p}{d p\over d\eta}=-{\cal H}-{\partial \phi\over \partial \eta}-\hat p^i{\partial \psi\over \partial x^i},
\label{pph}
\end{eqnarray}
where $p$ is the physical energy of the photon
(equivalent to the physical momentum in  units of $c=1$) 
in the cosmological rest frame, $\hat p^i$ is the unit vector of the photon momentum satisfying 
$\delta_{ij}\hat p^i \hat p^j=1$, we defined ${\cal H}=a'/a$, and the prime denotes differentiation with respect to the conformal time $\eta$. 

Using the relations 
\begin{eqnarray}
&&{d\psi(\eta,x^i(\eta))\over d \eta}={\partial \psi\over \partial \eta}
+{dx^i\over d\eta}{\partial \psi\over \partial x^i},
\\
&&{dx^i\over d\eta}=\hat p^i(1+\psi-\phi),
\end{eqnarray} 
and integrating from $\eta_j$ to $\eta_0$, 
where $\eta_j$ is the conformal time when the photon is emitted from
the $j$th object and we observe it at the present time $\eta_0$, we
find that Eq.~(\ref{pph}) yields the following solution up to the
first order of $\phi$ and $\psi$: 
\begin{eqnarray}
{p(\eta_0)\over p(\eta_j)}={a(\eta_j)\over a(\eta_0)}\exp\left\{-\int_{\eta_j}^{\eta_0}
\left({\phi'(\eta,\bm x(\eta))}-{\psi'(\eta,\bm x(\eta)}\right)d\eta
-\psi(\eta_0,\bm x(\eta_0))+\psi(\eta_j,\bm x(\eta_j))\right\},
\end{eqnarray}
where $\bm x(\eta_j)$ is the position of the $j$th object at the emission of the photon
(see Fig.~\ref{fig:configurations}). 
Hereafter, we assume $a(\eta_0)=1$. 
When the $j$th object has the peculiar velocity ${\bm v_j}$, the observational redshift is
multiplied by the factor
$({1+\bm \gamma\cdot {\bm v}_j)/\sqrt{1-\bm v_j^2}}$, 
where $\bm \gamma$ is the unit vector of the line-of-sight direction.  
Then, the redshift of the $j$th object is given by 
\begin{eqnarray}
1+z_j&=&
{1\over a(\eta_j)}\exp\left\{\int_{\eta_j}^{\eta_0}
\left({\phi'}-{\psi'}\right)d\eta
+\psi(\eta_0,\bm x(\eta_0))-\psi(\eta_j,\bm x(\eta_j))\right\}
{1+\bm \gamma\cdot {\bm v}_j\over \sqrt{1-\bm v_j^2}}.
\label{abc}
\end{eqnarray}

We rewrite the expression on the right-hand-side of Eq.~(\ref{abc}) as follows. 
First, assuming nonrelativistic motion ($|\bm v|\ll 1$), we may write
\begin{eqnarray}
{1+\bm \gamma\cdot {\bm v}_j\over \sqrt{1-\bm v_j^2}}
\simeq 1+\bm \gamma\cdot {\bm v}_j+{1\over 2}\bm v_j^2.
\end{eqnarray}
Here we call the terms $\bm \gamma\cdot {\bm v}_j$ and ${1\over 2}\bm v_j^2$
the first-order Doppler term and the second-order Doppler term, respectively. 
Up to the first order of metric perturbations, we also write
\begin{eqnarray}
&&\exp\left\{\int_{\eta_j}^{\eta_0}
\left({\phi'}-{\psi'}\right)d\eta
+\psi(\eta_0,\bm x(\eta_0))-\psi(\eta_j,\bm x(\eta_j))\right\}
\nonumber\\&&
~~~~~~~~
\simeq1+\int_{\eta_j}^{\eta_0}
\left({\phi'}-{\psi'}\right)d\eta
+\psi(\eta_0,\bm x(\eta_0))-\psi(\eta_j,\bm x(\eta_j)).
\end{eqnarray}
In this expression, we call the term $\int_{\eta_j}^{\eta_0}\left({\phi'}-{\psi'}\right)d\eta$
the integrated Sachs--Wolfe term. $\psi(\eta_0,\bm x(\eta_0))$ and $\psi(\eta_j,\bm x(\eta_j))$
represent the gravitational potential of the observer and that of the $j$th object, respectively. 

Furthermore, by introducing the time $\eta_1$ and setting  $\eta_j=\eta_1+\Delta\eta_j$
(see Fig.~\ref{fig:configurations}), we expand $1/a(\eta_j)$ as
\begin{eqnarray}
&&{1\over a(\eta_j)}\simeq{1\over a(\eta_1)}\biggl\{1-{\cal H}(\eta_1)\Delta\eta_j +
\left({\cal H}^2(\eta_1)-{1\over 2}{a''(\eta_1)\over a(\eta_1)}\right)\Delta\eta_j^2+{\cal O}(\Delta \eta_j^3)\biggr\}.
\label{ojoj}
\end{eqnarray}
In this expansion, as shown in Fig.~\ref{fig:configurations}, we assume that a photon is emitted 
from the object located at the position specified by the comoving distance $\chi=\eta_0-\eta_1$
and $\bm x_\perp=0$ at the conformal time $\eta_1$ and that the observer receives the photon at
the time $\eta_0$, where we suppose that the position is the centre of a cluster or a void.
We call this object the {\it reference object}. 
In Eq.~(\ref{ojoj}), we call the terms $-{\cal H}(\eta_1)\Delta \eta_j$ and
$({\cal H}^2(\eta_1)-a''(\eta_1)/2a(\eta_1))\Delta\eta_j^2$
the first-order Hubble term and the second-order Hubble term, respectively. 

Combining the above results, we have
\begin{eqnarray}
1+z_j&\simeq& (1+z_1)
\biggl\{1-{\cal H}(\eta_1)\Delta\eta_j +
\left({\cal H}^2(\eta_1)-{1\over 2}{a''(\eta_1)\over a(\eta_1)}\right)\Delta\eta_j^2
+\int_{\eta_j}^{\eta_0}\left({\phi'}-{\psi'}\right)d\eta
\nonumber\\
&&~~~~~~~
+\psi(\eta_0,\bm x(\eta_0))
-\psi(\eta_j,\bm x(\eta_j))+\bm \gamma\cdot {\bm v}_j+{1\over 2}|\bm v_j|^2
\biggr\},
\label{abcd}
\end{eqnarray} 
where we introduced $a(\eta_1)=1/(1+z_1)$. This is the redshift of the $j$th object.
\red{We neglected the Doppler effect of the peculiar motion of the observer because
we consider the relative redshift of objects located in a small region on the
well-subhorizon scales. }
The names of the terms in Eq.~(\ref{abcd})  are summarised in Table I. 
\begin{figure}[t]
\begin{center}
    \includegraphics[width=13cm]{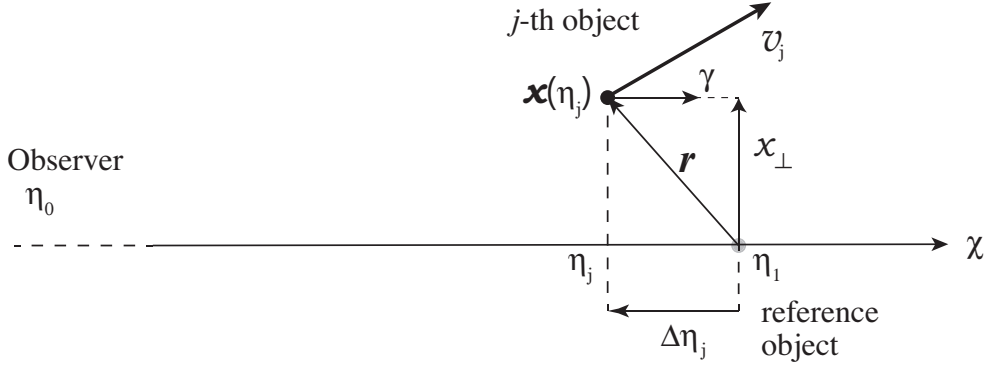}
\caption{Variables and coordinates. $\bm \gamma$ is the unit vector of the line-of-sight direction, and we also define $\chi=\eta_0-\eta$.
\label{fig:configurations}}
\end{center}
\end{figure}
\begin{table}[b]
\caption{Names of the terms in Eq.~(\ref{abcd}).
\label{tab:lrg_halodz}}
\begin{center}
\begin{tabular}{ll}
\hline
\hline
~~~~~~~~
$-(1+z_1){\cal H}(\eta_1)\Delta \eta_j$ & (first-order) Hubble term~~~~~~~~~~~
\\
~~~~~~~~
$(1+z_1)({\cal H}^2-a''/2a)|_{\eta=\eta_1}\Delta\eta_j^2$ ~~~~~& (second-order) Hubble term
\\
~~~~~~~~
$(1+z_1)\int_{\eta_j}^{\eta_0}({\phi'}-{\psi'})d\eta$ & integrated Sachs--Wolfe term
\\
~~~~~~~~
$-(1+z_1)\psi(\eta_j,\bm x(\eta_j))$ & gravitational potential term
\\
~~~~~~~~
$(1+z_1)\bm \gamma\cdot {\bm v}_j$ & (first-order) Doppler term
\\
~~~~~~~~
$(1+z_1){|\bm v_j|^2/2}$ & (second-order) Doppler term
\\
\hline
\hline
\end{tabular}
\end{center}
\end{table}
We also omit the integrated Sachs--Wolfe term and $\psi(\eta_0,\bm x(\eta_0))$ in Eq.~(\ref{abcd})
because we consider the relative redshift of well-subhorizon objects.
The gravitational potential of the observer $\psi(\eta_0,\bm x(\eta_0))$ does not
contribute to the final result of the relative redshift.
Then, in the present paper, we consider 
\begin{eqnarray}
1+z_j&\simeq& (1+z_1)
\biggl\{1-{\cal H}(\eta_1)\Delta\eta_j +
\left({\cal H}^2(\eta_1)-{1\over 2}{a''(\eta_1)\over a(\eta_1)}\right)\Delta\eta_j^2
-\psi(\eta_j,\bm x(\eta_j))+\bm \gamma\cdot {\bm v}_j+{1\over 2}|\bm v_j|^2
\biggr\}.
\label{abcde}
\end{eqnarray} 

The gravitational redshift of the reference object is expressed as
\begin{eqnarray}
1+z_r&\simeq& (1+z_1)
\biggl\{1-\psi(\eta_1,\bm x(\eta_1))
+\int_{\eta_1}^{\eta_0}\left({\phi'}-{\psi'}\right)d\eta
+\bm \gamma\cdot {\bm v}_1+{1\over 2}\bm v_1^2
\biggr\}.
\label{abcdee}
\end{eqnarray}
For example, the reference object is the central galaxy in a halo in Sec.~III. 
The difference between $z_j$ and $z_r$ is approximately given by
\begin{eqnarray}
\delta z_{jr}\equiv z_j-z_r&\simeq& (1+z_1)
\biggl\{
\psi(\eta_1,\bm x(\eta_1))-\psi(\eta_j,\bm x(\eta_j))
+\bm \gamma\cdot {\bm v}_j+{1\over 2}\bm v_j^2
-\bm \gamma\cdot {\bm v}_1-{1\over 2}\bm v_1^2
\biggr\}.
\label{abcdef}
\end{eqnarray}
In this expression, we omitted the Hubble term.
This is the basic formula adopted in Sec. III in the present paper (cf. Refs.~\cite{Kaiser,Cai}).
However, the reference object is not necessarily introduced in Secs. IV and V, where we use
Eq.~(\ref{abcde}).

\red{
  When we consider clusters of galaxies as in Secs. III and IV, we omit the Hubble term
  in Eq.~(\ref{abcde}).
  This omission is justified for clusters of galaxies because galaxy clusters do not
  expand. On the contrary, we include the Hubble term when we consider voids in Sec.~V.}

In the next sections we consider applications of this section. In these analyses, 
the observational quantities are obtained by averaging the redshifts of objects
over the velocity space with the phase space distribution function of objects
$f(\bm x, \bm v)$,
which is defined on the lightcone coordinate  \red{(see Figure \ref{fig:LC}).
The phase space distribution function $f(\bm x, \bm v)$ on the lightcone coordinate
is nontrivially related to the reference frame phase space distribution
function $f_{\rm RF}(t,\bm x,\bm v)$ by }
\begin{eqnarray}
f(\bm x, \bm v)=(1+\bm\gamma\cdot \bm v)f_{\rm RF}(t,\bm x,\bm v)\big|_{\rm ~light~cone}.
\label{RFdsf}
\end{eqnarray}
\red{The reason why  the phase space distribution function on the lightcone coordinate
  depends on the velocity in an asymmetric way with respect to $\bm\gamma\cdot\bm v$
  is understood with Figure \ref{fig:LC} (see the caption).}
Kaiser pointed out that this fact gives rise to an additional second-order Doppler
term $\langle (\bm \gamma\cdot \bm v)^2\rangle$ in the process of averaging the
redshifts over velocity space.
We further average the redshifts over spatial coordinates depending on the situation and the 
measurement strategy used for the gravitational redshift, 
where the spatial coordinates consist of the radial coordinate of the line-of-sight direction,
$\chi$, and the coordinates $\bm x_\perp$, perpendicular to $\chi$. 
In Sec.~III, we perform an average over $\chi$ and $\bm x_\perp$, while, 
in Secs.~IV and V, we perform an average only over $\chi$ in some ranges. 
In the present paper, for simplicity, we neglect the surface brightness modulation 
effect \cite{Kaiser,Jimeno}.

\begin{figure}[t]
\begin{center}
    \includegraphics[width=7cm]{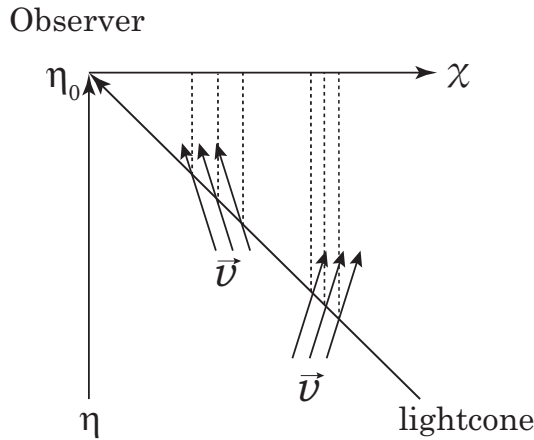}
    \caption{A sketch to explain the lightcone effect on the phase-space distribution function.
      Even when the density of objects in the reference frame is the same, the density defined
      on the lightcone coordinate $\chi$, which is obtained by projecting the points on which
      the objects intersect with the lightcone $\chi$, depend on the velocity
      $\bm v$ of the objects in an asymmetric way with respect to $\bm\gamma\cdot\bm v$.
\label{fig:LC}}
\end{center}
\end{figure}

\section{Satellite galaxies virialised in halos}
The authors of Refs.~\cite{Wojtak,Zhao,Jimeno} have reported measurements
of the gravitational redshift of galaxies in and near clusters relative to bright cluster galaxies,
which we revisit in this section. Here we restrict satellite galaxies virialised in a halo, and we 
develop a simple model of the gravitational redshift relative to central galaxies.
Our model is based on the halo approach, which is useful for describing the distribution of 
the dark matter as well as galaxies from large scales to 
small scales \cite{White,Seljak,CooraySheth2002}. 
Here we adopt the halo approach with central galaxies and satellite galaxies 
that fits the clustering of galaxies in redshift space
\cite{ReidSpergel,Zheng2005,HikageYamamoto2013,HikageYamamoto2015,HMTS,Kanemaru}.
We assume that the central galaxies are located 
at the centre of halos with negligible velocity dispersion and that the satellite 
galaxies are off-centred and moving with large virial random velocity.

Applying the result in the previous section, we regard the $j$th object as a 
satellite galaxy and the reference object as the central galaxy in a halo. 
Then, we take the following average of the redshift (\ref{abcdef}) over the 
satellite galaxies with the phase space distribution function $f(\bm x, \bm v)$:
\begin{eqnarray}
\langle \delta z\rangle = {{ \int d^2x_\perp\int d\chi \int d^3 v_j 
\delta z_{jr} f(\bm x,\bm v_j) }
\over { \int d^2x_\perp\int d\chi \int d^3 v_j f(\bm x,\bm v_j)}}
.
\label{deltazsatellite}
\end{eqnarray}
We take the effect of the lightcone coordinate by using Eq.~(\ref{RFdsf}),
and we assume $\int d\chi \int d^3 v_j (\bm \gamma\cdot \bm v_j)f_{\rm RF}(\bm x,\bm v_j)=0$ owing 
to the spherical symmetry of the system statistically.
Then, Eq.~(\ref{deltazsatellite}) leads to
\begin{eqnarray}
    \langle \delta z\rangle &=&  (1+z_1)\left(\psi(0) -\langle \psi_j \rangle
+\langle (\bm \gamma\cdot \bm v_j)^2 \rangle+{1\over 2}\langle |\bm v_j|^2 \rangle\right),
\end{eqnarray}
where $\psi(0)$ denotes the gravitational potential at the centre of a halo and 
$\langle \psi_j \rangle$ denotes the expectation value of the gravitational potential of
satellite galaxies and where $\langle (\bm \gamma\cdot \bm v_j)^2\rangle$ and $\langle |\bm v_j|^2\rangle$
denote the variance of the random velocity of satellite galaxies in one dimension and in three dimensions,
respectively. These quantities are defined under the condition that the mass of a halo is fixed.
Here we omitted the random velocity of the central galaxies by
assuming that it is negligibly small compared with that of the satellite galaxies. 
We also omitted the Hubble terms. Assuming isotropy of the random velocity of 
satellite galaxies, $\langle (\bm \gamma\cdot \bm v_j)^2 \rangle=\langle\bm v_j^2 \rangle/3$,
we have
\begin{eqnarray}
\langle \delta z\rangle &=& (1+z_1)\left(\psi(0)-\langle \psi_s \rangle
+{5\over 2}\langle (\bm \gamma\cdot \bm v_j)^2 \rangle\right).
\end{eqnarray}
When we take the random velocity of the central galaxies into account,  $\langle (\bm \gamma\cdot \bm v_j)^2\rangle$ is
replaced by $\langle (\bm \gamma\cdot \bm v_j)^2\rangle-\langle (\bm \gamma\cdot \bm v_1)^2\rangle$,
where $\langle (\bm \gamma\cdot \bm v_1)^2\rangle$
is the one-dimensional velocity variance of a central galaxy.

For the one-dimensional random velocity variance of satellite galaxies, we here adopt the simplest
model, following Refs.~\cite{HMTS,HikageYamamoto2013,Kanemaru}: 
\begin{eqnarray}
\langle (\bm \gamma\cdot \bm v_j)^2\rangle=\sigma_{\rm v,off}^2(M_{\rm vir})={GM_{\rm vir}\over 2R_{\rm vir}},
\end{eqnarray}
where $r_{\rm vir}$ and $M_{\rm vir}$ are the virial radius and the virial mass, respectively. 
We assume that the density profile of halos follows the NFW profile \cite{NFW} 
\begin{eqnarray}
\rho_{\rm NFW}(r)={\rho_s\over (r/r_s)(1+r/r_s)^2}, 
\label{NFWprofile}
\end{eqnarray}
where $\rho_s$ and $r_s$ are the parameters.  
The gravitational potential $\psi$ follows the Poisson equation 
\begin{eqnarray}
\triangle \psi(r)=4\pi G \rho_{\rm NFW}(r),
\end{eqnarray}
which leads to the solution 
\begin{eqnarray}
\psi(r)=-{4\pi G\rho_sr_s^2}{\ln (1+r/r_s)\over (r/r_s)}. ~
\label{psiNFW}
\end{eqnarray}
We also assume that the satellite galaxy number density is proportional to  the NFW profile.
In this case, we may write
\begin{eqnarray}
\langle\psi_s\rangle
&=&4\pi\int_0^{r_{\rm vir}} drr^2 {\rho_{\rm NFW}(r)\over M_{\rm vir}} \psi(r)
\nonumber\\
&=&{4\pi r_s^3\rho_s\over M_{\rm vir}}\left(-{4\pi G\rho_sr_s^2}\right)\int_0^c dx{\ln(1+x)\over (1+x)^2}
\nonumber\\
&=&-{c\over m^2(c)}{c-\log(1+c)\over 1+c}{GM_{\rm vir}\over r_{\rm vir}}.
\end{eqnarray}
We introduce the concentration parameter $c$ and the virial mass $M_{\rm vir}$ instead 
of $\rho_{\rm s}$ and $r_{\rm s}$, by $c={r_{\rm vir}}/{r_{\rm s}}$ and 
$M_{\rm vir}=M(<r_{\rm vir})={4\pi}r_{\rm vir}^3\Delta_{\rm vir}\bar \rho_{\rm m}/3$, where 
$M(<r_{\rm vir})$ is the mass within the radius $r_{\rm vir}$, $\bar \rho_{\rm m}$ is 
the mean matter density, and $\Delta_{\rm vir}$ is the density contrast 
of a halo, for which we here adopt $\Delta_{\rm vir}=265$ at $z=0.3$. 
From Eq.~(\ref{psiNFW}) we have
\begin{eqnarray}
\psi(0)=-{4\pi G\rho_sr_s^2}=-{GM_{\rm vir}\over r_{\rm vir}}{c\over m(c)}.
\end{eqnarray}
where we defined $m(c)=\ln(1+c)-c/(1+c)$, and 
\begin{eqnarray}
\psi(0)-\langle\psi_s\rangle={GM_{\rm vir}\over r_{\rm vir}}
\left(-{c\over m(c)}+{c\over m^2(c)}{(c-\log(1+c))\over 1+c}\right).
\end{eqnarray}

Combining the results, we have
\begin{eqnarray}
\langle \delta z\rangle&=&(1+z_1)\left(\psi(0)-\langle\psi_s\rangle+
\langle (\bm \gamma\cdot \bm v_j)^2\rangle\right)
\nonumber\\
&=&
(1+z_1){GM_{\rm vir}\over r_{\rm vir}}
\left(-{c\over m(c)}+{c\over m^2(c)}{(c-\log(1+c))\over 1+c}+{5\over 4}\right).
\label{deltazMM}
\end{eqnarray}

\red{In the latter part of this section, we present a theoretical prediction with the halo
  occupation distribution with central galaxy and satellite galaxy.}
Note that $\langle \delta z\rangle$ of Eq.~(\ref{deltazMM}) is a 
function of the halo mass $M_{\rm vir}$, which we write as $\langle \delta z(M_{\rm vir })\rangle$. 
Hereafter we write $M_{\rm vir}$ as $M$. In a practical analysis, we need 
to use a large number of satellite galaxies in different halo masses.
We investigate the theoretical expectation value of the gravitational redshift
by integrating over the halo mass function with the HOD
for satellite galaxies.
%
We follow the HOD fitting functions  
for the central galaxies and the satellite galaxies
proposed in Ref.~\cite{Zheng2005}:
\begin{eqnarray}
&&N_{\rm HOD}(M)=\langle N_{\rm cen}\rangle(1+\langle N_{\rm sat}\rangle), \\
&&\langle N_{\rm cen}\rangle =\frac{1}{2}\left[1+{\rm erf}\left(\frac{\log_{10}(M)-\log_{10}
(M_{\rm min})}{\sigma_{\log M}}\right)\right], \\
&&\langle N_{\rm sat}\rangle =
\left(\frac{M-M_{\rm cut}}{M_1}\right)^{\alpha},
\label{eq:HOD}
\end{eqnarray}
where ${\rm erf}(x)$ is the error function.

Table \ref{tab:lrg_halo} lists the HOD parameters for the three
galaxy samples, the luminous red galaxy (LRG) sample of the Sloan 
Digital Sky Survey (SDSS) II \cite{ReidSpergel}, 
the low redshift (LOWZ) sample \cite{Parejko}, and the CMASS sample \cite{Manera} of 
the Baryon Oscillation Spectroscopy Survey (BOSS) SDSS III. We assume
that the mean redshifts of these galaxy samples are $z_{\rm mean}=0.32$ for the LRG sample
and the LOWZ sample and $z_{\rm mean}=0.56$ for the CMASS sample.

\begin{table}[t]
\caption{HOD parameters and the mean redshift for the LRG samples~\cite{ReidSpergel},
 the LOWZ sample \cite{Parejko}, 
and the CMASS (mock) sample \cite{Manera}. The mean redshift is used as the value of $z_1$.
\label{tab:lrg_halo}
}
\begin{center}
\begin{tabular}{cccc}
\hline
\hline
~ & ~~~~~~~~~~~~~LRG~~~~~~~~~~~~~& ~~~~~~~~~~~~~LOWZ ~~~~~~~~~~~~~& ~~~~~~~~~~~~~CMASS ~~~~~~~~~~~~~ \\
\hline
~~~~~
$M_{\rm min}$ & $5.7\times 10^{13}\;h^{-1}M_\odot$ 
& $1.5\times 10^{13}\;h^{-1}M_\odot$
& $1.2\times 10^{13}\;h^{-1}M_\odot$ 
\\
~~~~~
$\sigma_{\log M}$ & 0.7 & 0.45 & 0.596 
\\
~~~~~
$M_{\rm  cut}$ & $3.5\times 10^{13}\;h^{-1}M_\odot$  
&  $1.4\times 10^{13}\;h^{-1}M_\odot$
&  $1.2\times 10^{13}\;h^{-1}M_\odot$ \\
~~~~~
$M_1$ & $3.5\times10^{14}\;h^{-1}M_\odot$ 
& $1.3\times10^{14}\;h^{-1}M_\odot$ 
& $1.0\times10^{14}\;h^{-1}M_\odot$  \\
~~~~~
$\alpha$ & $1$ & $1.38$ & $1.0$ \\
~~~~~
  {\rm mean redshift }
& $0.32$ & $0.32$ & $0.56$\\
\hline\hline
\end{tabular}
\end{center}
\vspace{0.7cm}
\caption{Contributions of the gravitational potential $(1+z_1)(\psi(0)-\langle\psi_s\rangle)$, 
  the Doppler term $(1+z_1){5\over2}\langle (\bm \gamma\cdot \bm v)^2\rangle$, and the
  total $\langle\delta z\rangle$ for the LRG sample, the LOWZ sample, and the CMASS sample.
Note that $\bar r_{\rm vir}$ shows the averaged virial radius for each galaxy sample. 
\label{tab:lrg_halodz}}
\begin{center}
\begin{tabular}{cccc}
\hline
\hline~ ~~~~~~~~~~~~~~& ~~~~~~~~~LRG~~~~~~~~~& ~~~~~~~~~LOWZ~~~~~~~~~ & ~~~~~~~~~CMASS~~~~~~~~~\\
\hline
~~
$(1+z_1)(\psi(0)-\langle\psi_s\rangle)$ 
~~~~~
  & $-3.2 \times 10^{-5}\; (-9.7\;{\rm km/s})$
~~~~~
 & $-2.3 \times 10^{-5}\; (-7.0\;{\rm km/s})$ 
~~~~~
  & $-1.8 \times 10^{-5}\; (-5.4\;{\rm km/s})$ \\
~~~~~
$(1+z_1){5\over2}\langle (\bm \gamma\cdot \bm v)^2\rangle$ 
~~~~~
  & $+1.7\times 10^{-5}\; (+5.1\;{\rm km/s})$
~~~~~
 & $+1.2 \times 10^{-5}\; (+3.5\;{\rm km/s})$ 
~~~~~
  & $+1.0\times 10^{-5}\; (+3.0\;{\rm km/s})$ \\
~~~~~
$\langle\delta z\rangle$ 
~~~~~\;
  & $-1.5 \times 10^{-5}\; (-4.6{\rm km/s})$
~~~~~
 & $-1.2 \times 10^{-5}\; (-3.5\;{\rm km/s})$ 
~~~~~
  & $-0.8 \times 10^{-5}\; (-2.5\;{\rm km/s})$ \\
${\bar r_{\rm vir}}$ &$1.0~h^{-1}{\rm Mpc}$&$0.85~h^{-1}{\rm Mpc}$& $0.79~h^{-1}{\rm Mpc}$\\
\hline\hline
\end{tabular}
\end{center}
\vspace{0.7cm}
\caption{Same as Table \ref{tab:lrg_halodz}, but for the modified gravity case 
  with the effective gravitational constant $G_{\rm eff}={4G/3}$,
  which only modifies the velocity dispersion of satellite galaxies. 
\label{tab:lrg_halodzmg}
}
\begin{center}
\begin{tabular}{cccc}
\hline
\hline~ ~~~~~~~~~~~~~~& ~~~~~~~~~LRG~~~~~~~~~& ~~~~~~~~~LOWZ~~~~~~~~~ & ~~~~~~~~~CMASS~~~~~~~~~\\
\hline
~~
$(1+z_1)(\psi(0)-\langle\psi_s\rangle)$ 
~~~~~
  & $-3.2 \times 10^{-5}\; (-9.7\;{\rm km/s})$
~~~~~
 & $-2.3 \times 10^{-5}\; (-7.0\;{\rm km/s})$ 
~~~~~
  & $-1.8 \times 10^{-5}\; (-5.4\;{\rm km/s})$ \\
~~~~~
$(1+z_1){5\over2}\langle (\vec \gamma\cdot \vec v)^2\rangle$ 
~~~~~
  & $+2.3\times 10^{-5}\; (+6.8\;{\rm km/s})$
~~~~~
 & $+1.6 \times 10^{-5}\; (+4.7\;{\rm km/s})$ 
~~~~
  & $+1.3\times 10^{-5}\; (+3.9\;{\rm km/s})$ \\
~~~~~
$\langle\delta z\rangle$ 
~~~~~
  & $-1.0 \times 10^{-5}\; (-3.0\;{\rm km/s})$
~~~~~
 & $-0.8 \times 10^{-5}\; (-2.3\;{\rm km/s})$ 
~~~~~
  & $-0.5 \times 10^{-5}\; (-1.5\;{\rm km/s})$ \\
\hline\hline
\end{tabular}
\end{center}
\vspace{0.7cm}
\caption{Same as  Table \ref{tab:lrg_halodz}, but for the case in which the
  central galaxies have a random velocity with $30$\% of that of the satellite galaxies.
\label{tab:lrg_halodzmgd}
}
\begin{center}
\begin{tabular}{cccc}
\hline
\hline~ ~~~~~~~~~~~~~~& ~~~~~~~~~LRG~~~~~~~~~& ~~~~~~~~~LOWZ~~~~~~~~~ & ~~~~~~~~~CMASS~~~~~~~~~\\
\hline
~~
$(1+z_1)(\psi(0)-\langle\psi_s\rangle)$ 
~~~~~
& $-3.2 \times 10^{-5}\; (-9.7\;{\rm km/s})$
~~~~~
 & $-2.3 \times 10^{-5}\; (-7.0\;{\rm km/s})$ 
~~~~~
  & $-1.8 \times 10^{-5}\; (-5.4\;{\rm km/s})$ \\
~~~~~
$(1+z_1){5\over2}\langle (\vec \gamma\cdot \vec v)^2\rangle$ 
~~~~~
  & $+1.5\times 10^{-5}\; (+4.6\;{\rm km/s})$
~~~~~
 & $+1.1 \times 10^{-5}\; (+3.2\;{\rm km/s})$ 
~~~~
  & $+0.9\times 10^{-5}\; (+2.7\;{\rm km/s})$ \\
~~~~~
$\langle\delta z\rangle$ 
~~~~~
  & $-1.7 \times 10^{-5}\; (-5.1\;{\rm km/s})$
~~~~~
 & $-1.3 \times 10^{-5}\; (-3.8\;{\rm km/s})$ 
~~~~~
  & $-0.9 \times 10^{-5}\; (-2.8\;{\rm km/s})$ \\
\hline\hline
\end{tabular}
\end{center}
\end{table}

The halo mass function $dn/dM$ is the number density of 
halos with  mass $M$ per unit volume and per unit mass.
Namely, the halo mass function $(dn/dM)dM$ describes the comoving number density of 
halos of the mass in the range $M\sim M+dM$.
A fitting function of the halo mass function has been investigated with numerical
simulations by several authors, fitted in the form \red{\cite{ShethTormen1999,ST2,Parkinson}}
\begin{eqnarray}
 M{dn\over dM}={\bar \rho_m \over M}{d\ln \sigma_R^{-1}\over d\ln M} f(\sigma_R),
\end{eqnarray}
where $\sigma_R$ is the root-mean-square fluctuation in spheres containing
 mass $M$ at the initial time, which is extrapolated to redshift $z$ using linear theory, 
$\delta_c$ is the critical value of the initial overdensity 
that is required for collapse, and $\delta_c=1.69$ is adopted.
In the present paper, we adopt the fitting formula in Ref.~\cite{ShethTormen1999}:
\begin{eqnarray}
 f(\sigma_R)=0.322\sqrt{2\times0.707\over \pi}\left[
1+\left({\sigma_R\over 0.707\delta_c}\right)^{0.3}\right]{\delta_c\over \sigma_R}\exp
\left(-{0.707\delta_c^2\over 2\sigma_R^2}\right).
\end{eqnarray}

We compute the ensemble average of the gravitational redshift 
over the halo mass function with the satellite galaxy HOD by using
\begin{eqnarray}
\langle\delta z\rangle={{\int dM {dn\over dM} \langle N_{\rm sat}\rangle \delta z(M)}
\over {\int dM {dn\over dM} \langle N_{\rm sat}\rangle} },
\label{MMNN}
\end{eqnarray}
where we insert Eq.~(\ref{deltazMM}) into the right-hand-side of Eq.~(\ref{MMNN}).
Table \ref{tab:lrg_halodz} lists the result of 
$\langle\delta z\rangle$, which is the combination of 
the gravitational potential term $(1+z_1)(\psi(0)-\langle\psi_s\rangle)$ 
and the second-order Doppler terms 
$(1+z_1){5\over2}\langle(\bm\gamma\cdot \bm v)^2\rangle$, for each galaxy sample. 
The result shows that the gravitational potential term $(1+z_1)(\psi(0)-\langle\psi_s\rangle)$
is the order of ${\rm a~few}\times 10^{-5}$, while the second-order Doppler 
term $\langle(\bm\gamma\cdot \bm v)^2\rangle$ makes a significant contribution, and the
total amplitude of the signal is significantly reduced.
Because we consider the satellite galaxies virialised in halos, the typical 
separation between the central galaxy and the satellite is the virial radius 
of the order of $1~h^{-1}$~Mpc, where the second-order Doppler term 
makes a significant contribution to the total amplitude of the signal.
In the velocity unit, the amplitude of the signal is from $5$ to $2$ km/s depending on the samples.
This is consistent with previous results \cite{Wojtak,Zhao,Jimeno}.
The result also shows that the amplitude of the signal decreases from left to right in  Table III.
This HOD dependence is understood as the change of an averaged size of halos for each galaxy sample.
The averaged virial radius $\bar r_{\rm vir}$, which is defined similarly to Eq.~(\ref{MMNN}),  
decreases from left to right in  Table III, 
which means that the gravitational potential of satellite galaxies becomes shallower accordingly. 

An interesting application of measurements of the gravitational redshift is the testing modified gravity
models. For example, in an $F(R)$ gravity model, when the screening mechanism does not work, the
velocity of satellite galaxies increases owing to the scalar force. However, the gravitational potential
does not changes as long as the matter density profile is the same. When the effective gravitational
constant as well as the variance of the random velocity increases by the factor $4/3$, the signal of
the gravitational redshift is evaluated as indicated in Table \ref{tab:lrg_halodzmg}.
This demonstrates that the gravitational redshift is potentially an interesting test of modified gravity \cite{Zhao}.
However, the random velocity of central galaxies could be a systematic error \cite{Wojtak,Zhao,Jimeno}.
When the central galaxy has a random velocity dispersion of $30$\%
of the satellite galaxies, the prediction changes, as shown in Table \ref{tab:lrg_halodzmgd}.
Thus, we need further investigations of the errors and systematics of the method as a test of gravity 
theories. 

  We simply estimate the error of the averaged central-satellite velocity
  difference $\Delta v^{\rm cen-sat}$ in a given galaxy samples by inverse-variance weighted averaging as follows:
\begin{equation}
(\Delta v^{\rm cen-sat})^{-2}=V \int dM\frac{dn}{dM}\langle N_{\rm cen}\rangle\langle N_{\rm sat}\rangle \sigma_{\rm vir}^{-2} ,
\end{equation}
where $V$ is the survey volume, $\sigma_{vir}$ is the Virial velocity
of the host halo with mass $M$ and we set $V=1.6(h^{-1}{\rm Gpc})^2$
for SDSS LRG, $V=0.79(h^{-1}{\rm Gpc})^2$ for LOWZ, and $V=1.75(h^{-1}{\rm
  Gpc})^2$ for CMASS samples. We find that $\Delta v^{\rm cen-sat}$
becomes 5.0 km/s (SDSS LRG), 2.7 km/s (LOWZ), and 1.4 km/s (CMASS). The
signal-to-noise ratio of $\delta z$ is 1--2 in the current sample.  The
statistical error will be improved in future galaxy surveys such as
DESI, PFS, and Euclid, which cover a larger survey volume. Here we assume
that all of the central galaxies are identified. The gravitational
redshift signal weakens depending on the fraction of misidentified
central galaxies.

\section{Intracluster gas}
The recent studies of the gravitational redshift of clusters galaxies focused on measurements
of galaxies \cite{Wojtak,Zhao,Jimeno,Kaiser,Cai}.
Motivated by the recent X-ray observations by the Hitomi satellite, in which
intracluster gas motions were investigated with an accuracy of the order of $10$ to $20$ km/s,
we next consider the gravitational redshift of intracluster gas. 
The gravitational redshift of X-ray gas was investigated in Ref.~\cite{BroadHurst}, but we here
explain another motivation for considering this problem.
Nonthermal pressure of the intracluster gas is an unsolved problem in cluster physics.
From  cosmological hydrodynamical simulations, it is shown that intracluster gas motions can be
generated in the structure formation process and that nonthermal random motions contribute to the nonthermal
pressure (e.g., \cite{Battaglia,Shaw}).
This nonthermal pressure might cause a discrepancy between the hydro-equilibrium mass and the lensing mass.
In this section, we assume that small-scale random motions of intracluster gas cause the nonthermal pressure,
and we investigate a possible signal of the gravitational redshift based on a simple model of the intracluster gas
including the nonthermal pressure.

We start by assuming that intracluster gas particles follow the Boltzmann distribution function (e.g., \cite{YSS})
\begin{eqnarray}
  f_{\rm RF}({\bm x},{\bm v})=n({\bm x})\left({2\pi \over mT({\bm x})}\right)^{3/2}
  \exp\left\{-m{[{\bm v}-{\bm V}({\bm x})]^2\over 2T({\bm x})}\right\},
\end{eqnarray}
which is characterised by particle number density $n({\bm x})$ of a specific element
at position $\bm x$,
temperature $T(\bm x)$, the peculiar velocity field of the random motions, ${\bm V}({\bm x})$, and the
mass of the particle, $m$, where we consider iron particles. 
Here we adopt units in which the Boltzmann constant equals one.

From Eq. (\ref{abcde}), omitting the Hubble term, we can write the redshift of a particle as
\begin{eqnarray}
1+z_{j} 
&\simeq& 1+z_1+(1+z_1)\biggl\{-\psi(\eta_j,\bm x(\eta_j))+\bm \gamma\cdot {\bm v}_j+{1\over 2}\bm v_j^2\biggr\}.
\label{abcdefg}
\end{eqnarray}
Then, we define the gravitational redshift projected along the line-of-sight direction
by integrating Eq.~(\ref{abcdefg}) over  velocity space and the line-of-sight coordinate:
\begin{eqnarray}
1+\langle z(x_\perp) \rangle&=&1+z_1
+ (1+z_1){\int d\chi \int d^3 v f(\bm x, \bm v)(\bm \gamma\cdot\bm v+{1\over 2}\bm v^2
-\psi(\bm x))
            \over
  \int d\chi \int d^3 v f(\bm x, \bm v)
}.
\end{eqnarray}
After integration with respect to the velocity, we have
\begin{eqnarray}
&&1+  \langle  z(x_\perp) \rangle=1+z_1
  +(1+z_1)
  {\int d\chi ~n(\bm x)\left\{\bm \gamma\cdot \bm V(\bm x)+(\bm \gamma\cdot \bm V(\bm x))^2
+{1\over 2}|\bm V(\bm x)|^2+{5\over 2}{T(\bm x)\over m}-\psi(\bm x)\right\}
\over
\int d\chi~ n(\bm x)(1+\bm\gamma\cdot\bm V)
  }.
\label{vvtp}
\end{eqnarray}
If we assume that the system is spherically symmetric, we may omit the linear term
$\int d\chi n(\bm x)\bm\gamma\cdot \bm V=0$.
This assumption will not be justified when spherical symmetry of the system is not
guaranteed. However, we may assume this spherical symmetry statistically when many clusters
are observed. Then, we have
\begin{eqnarray}
1+\langle  z_j(x_\perp) \rangle&=&1+z_1+(1+z_1)
{\int d\chi ~n(\bm x) \left((\bm \gamma\cdot \bm V(\bm x))^2
+{1\over 2}|\bm V(\bm x)|^2
+{5\over 2}{T(\bm x)\over m}-\psi(\bm x)\right)
              \over
              \int d\chi ~n(\bm x) }.
\label{vtp}
\end{eqnarray}
Here we assume isotropy of the peculiar velocity dispersion, 
$3\langle (\bm \gamma\cdot \bm V)^2 \rangle=\langle |\bm V|^2\rangle=\sigma_{\rm rnd}^2$,
where $\sigma_{\rm rnd}^2$ denotes the variance of the random motions of the gas.   
Furthermore, by assuming that the emissivity of the photon line emission is 
proportional to the number density 
of particles, i.e., the mass density of gas particles, $\rho_{\rm gas}$, 
Eq.~(\ref{vtp}) leads to
\begin{eqnarray}
1+\langle z(x_\perp) \rangle&=&
         1+z_1+    (1+z_1) {\int d\chi \rho_{\rm gas}(\bm x) \left({5\over 6}\sigma_{\rm rnd}^2(\bm x)
+{5\over 2}{T(\bm x)\over m}-\psi(\bm x)\right)
              \over
               \int d\chi \rho_{\rm gas}(\bm x) }.
\label{dz}
\end{eqnarray}
Note that the right-hand side of Eq.~(\ref{dz}) is a function of the projected radius $\chi_\perp$, and we
define the relative gravitational redshift by
\begin{eqnarray}
\langle \delta z(x_\perp) \rangle&=&
\langle z(x_\perp) \rangle-\langle z(0) \rangle.
\label{dz1}
\end{eqnarray}

When spherical symmetry of the system is guaranteed statistically, the integration of
the term $\bm \gamma \cdot \bm V(\bm x)$ in Eq.~(\ref{vvtp}) becomes zero; otherwise,
the term makes a large contribution. To estimate the variance of
this term, for simplicity, we estimate the variance of $\delta z$ by
\begin{eqnarray}
  \langle\delta z^2(x_\perp)\rangle&=&
  (1+z_1)^2\left\langle\left(
{\int d\chi n(\bm x) (\bm \gamma \cdot \bm V)
\over \int d\chi n(\bm x)}\right)^2
\right\rangle
\nonumber\\
  &\simeq&(1+z_1)^2
{\int d\chi n^2(\bm x) (\bm \gamma \cdot \bm V)^2
\over \int d\chi n^2(\bm x)}
=(1+z_1)^2{\int d\chi \rho_{\rm gas}^2(\bm x) {1\over 3}\sigma_{\rm rnd}^2
\over \int d\chi \rho_{\rm gas}^2(\bm x)},
\label{dz2}
\end{eqnarray}
where $\langle\cdot\rangle$ in the above equation means the ensemble average with
respect to the random motions $\bm V(\bm x)$.  

\begin{figure}[t]
\begin{center}
\includegraphics[width=180mm]{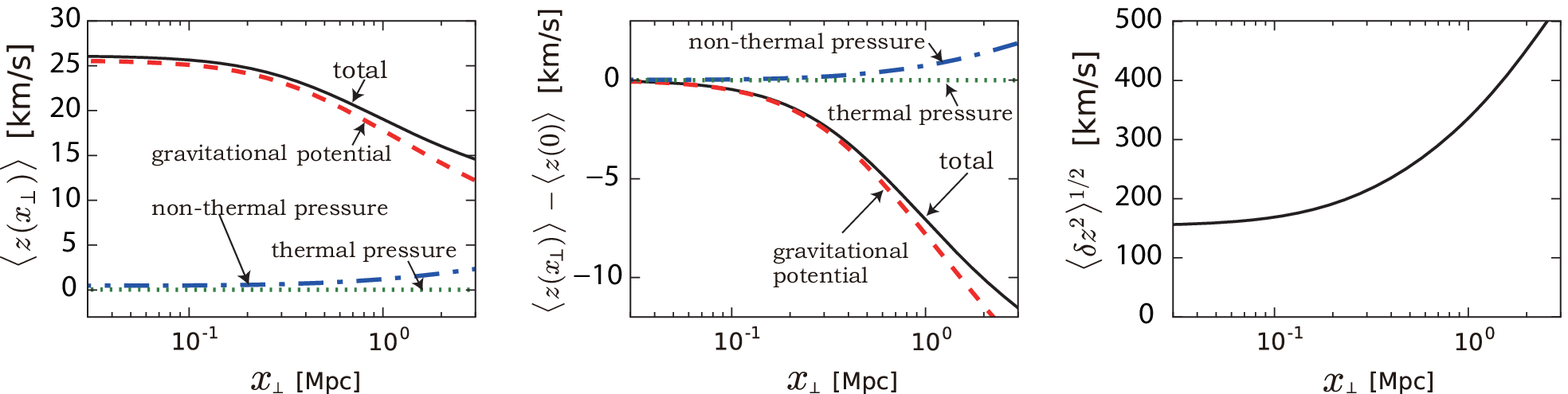}
 \caption{
Contribution to the gravitational redshift by
adopting a model for the Coma Cluster (see the Appendix for details).
 {\it Left panel:}  
The black curve shows the total of $\langle z(x_\perp)\rangle$ of Eq.~(\ref{dz}), 
 the red dashed curve shows the gravitational potential term $\psi(\bm x)$,
 the blue dash-dotted curve shows the nonthermal pressure term $5\sigma_{\rm rnd}^2/6$,
 and the green dotted curve shows the thermal pressure term $5T(\bm x)/2m$.
 {\it Middle panel:} The same as the left panel but for $\langle \delta z(x_\perp)
  \rangle =\langle z(x_\perp)\rangle-\langle z(0)\rangle$.  
 {\it Right panel:} The variance $\sqrt{\langle\delta z^2(x_\perp)
  \rangle}$ of Eq.~(\ref{dz2}). In this figure, we adopted $z_1=0$.
\label{fig:GR}}
\end{center}

\begin{center}
\vspace{0cm}
\includegraphics[width=180mm]{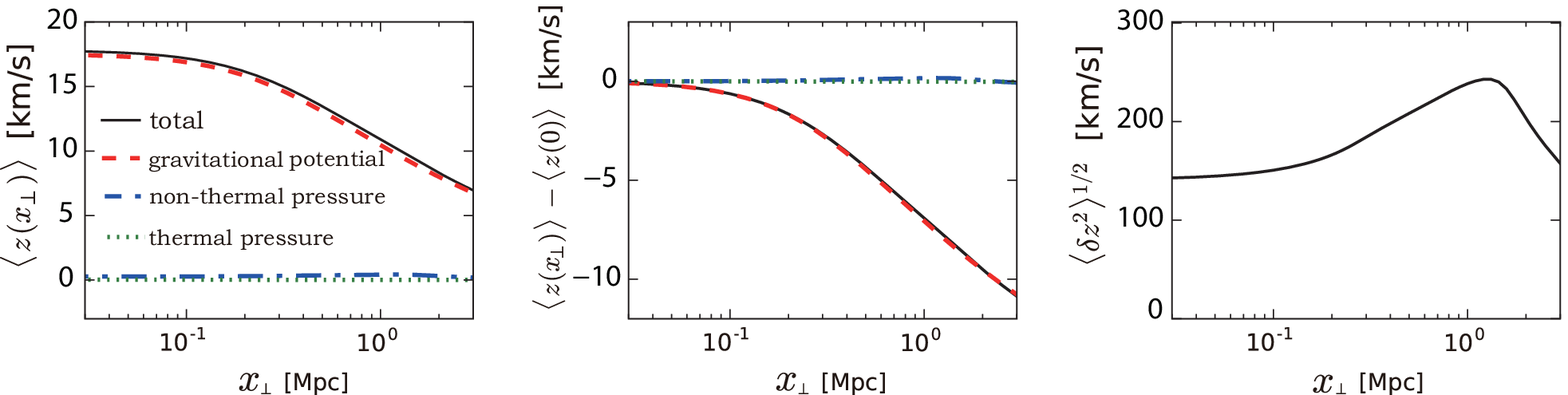}
\vspace{0cm}
\caption{Same as Fig. \ref{fig:GR}, but for the Perseus Cluster; 
the modelling parameters are from Urban et al. \cite{Urban} 
and Simionescu et al. \cite{Simionescu} (see also the Appendix).
\label{fig:GR_per}}
\end{center}
\end{figure}
\red{In the latter part of this section, we demonstrate the contribution of the terms of Eq.~(\ref{dz})
  adopting a simple model of intracluster gas.} The nonthermal pressure in intracluster gas is under debate
  based on cosmological hydrodynamical
simulations (e.g., \cite{Battaglia,Shaw}).
Their simulation results indicate that small-scale random motions of intracluster gas are generated
during the process of the cluster formation, which causes nonthermal pressure of the
intracluster gas. We therefore assume that the small-scale random motions are responsible
for the nonthermal pressure.
Following this scenario, we adopt the following model of random velocity 
$\sigma_{\rm rnd}^2(r)$ in connection with nonthermal pressure $P_{\rm nonthermal}(r)$  \cite{Battaglia,Shaw,Telkina1,Telkina2,Shi}: 
\begin{eqnarray}
  \rho_{\rm gas}(r)\sigma_{\rm rnd}^2(r)=P_{\rm nonthermal}(r).
\end{eqnarray}
The nonthermal pressure is estimated by the fraction $g(r)$ of the total pressure:
\begin{eqnarray}
  P_{\rm nonthermal}(r)=g(r)P_{\rm total}(r).
\end{eqnarray}
Hence, using $P_{\rm total}= g^{-1}P_{\rm nonthermal} = (1-g)^{-1}P_{\rm thermal}$,
we may write
\begin{eqnarray}
  P_{\rm nonthermal}(r) = {g(r) \over 1-g(r)}n_{\rm gas}(r)T(r),
\end{eqnarray}
where we used $P_{\rm thermal}(r)=n_{\rm gas}(r)T(r)$, and $n_{\rm gas}(r)$
is the number density of the particles in the intracluster gas. 
According to hydrodynamical simulations \cite{Battaglia,Shaw},
the nonthermal pressure component to the total pressure can
be modelled with the expression
\begin{eqnarray}
  g(r) = \alpha_{nt}(1 + z)^{\beta_{nt}}\left({r\over r_{500}}\right)^{n_{nt}}
 \left({M_{200}}\over 3\times 10^{14}\;M_{\odot}\right)^{n_M},
\end{eqnarray}
where $\alpha_{nt}$, $\beta_{nt}$, $n_{nt}$, and $n_M$ are constants.
Here $r_{500}$ and $M_{200}$ mean the radius and mass at the radius where the matter density 
in the galaxy cluster is $500$ and $200$ times of the critical density, respectively.
For our demonstration of the effects from the nonthermal pressure contribution, we adopt the parameter
values $(\alpha_{nt}, \beta_{nt}, n_{nt}, n_M) = (0.18, 0.5, 0.8, 0.2)$,
which are the best-fit values of the numerical simulations in Ref.~\cite{Shaw}.
The value of $\alpha_{nt}$ determines the contribution from the random motions 
of gas, $\sigma_{\rm rnd}^2$.

Figures \ref{fig:GR} and \ref{fig:GR_per} exemplify the behaviour of 
$\langle z(x_\perp)\rangle$ (left panel) and $\langle \delta z(x_\perp)\rangle$ (centre panel)
as a function of $x_\perp$ on the basis of theoretical models for the Coma 
Cluster and the Perseus Cluster, respectively, which are constructed to fit observations. 
The details of the theoretical models are summarised in the Appendix. 
In the left and centre panels of these figures, the red dashed curve is the gravitational potential
contribution, the blue dash-dotted curve is the nonthermal pressure term contribution,
and the green dotted curve is the thermal pressure term contribution.
The black solid curve is the total pressure. 
Thus, the amplitude of the relative gravitational redshift (centre panel) is of the order of 5--10 km/s. 
The gravitational potential term makes a dominant contribution to the gravitational redshift,
though the nonthermal pressure term makes a slight contribution.
The contribution from the thermal pressure is completely negligible. 
However, measurements of the outskirt region is necessary for detecting the signal of the relative gravitational redshift.

The right panels of Figs. \ref{fig:GR} and \ref{fig:GR_per} show
$\langle\delta z^2(x_\perp)\rangle^{1/2}$, which can be interpreted as
the dispersion of the signal in the $x_\perp$ direction. 
When the random motions of gas have coherent large-scale structures in a
halo, many clusters will be necessary to reduce the statistical errors
for the measurement of the gravitational redshift.
The error estimation will depend on the properties of the random motions of gas,
which is beyond scope of the present paper. 

\section{Void model} 
We next consider a possible signal of the gravitational redshift in measurements of galaxies
associated with voids. Voids are characteristic structures of the large-scale structure 
in the cold dark matter model universe. Recently, voids have become a useful tool for testing
cosmological models and gravity theories (e.g., see Refs.~\cite{UVP,Mao,Micheletti,CaiII,VIMOSP,Nico}).

In general, the region inside a void is not always completely empty, and some galaxies might
be found inside voids. This gives us a chance to find a possible signal of the gravitational
redshift of voids.
However, in the case of voids, in contrast to the case of clusters of galaxies, 
a galaxy is not always found at the centre of a void. 
Then, as in the case of the previous section, we consider the projection along the 
line-of-sight direction, and we consider the relative gravitational redshift as a function
of the projected radius, the coordinate perpendicular to the line-of-sight direction.
We consider the average 
\begin{eqnarray}
1+\langle z(x_\perp) \rangle &=& {{ \int d\chi \int d^3 v_j 
(1+z_j) f(\bm x,\bm v_j) }
\over {\int d\chi \int d^3 v_j f(\bm x,\bm v_j) }}
\nonumber\\
&=& 1+z_1+(1+z_1){{ \int d\chi n_{\rm g}(\chi,x_\perp)
\widetilde\delta z(\chi,x_\perp) }
\over {\int d\chi n_{\rm g}(\chi,x_\perp)}(1+\bm \gamma \cdot \bm V)},
\label{deltazvoid}
\end{eqnarray}
where $n_{\rm g}(\chi,x_\perp)$ is the galaxy number density, and we defined 
\begin{eqnarray}
\widetilde\delta z(\chi,x_\perp)&=&
\biggl\{-{\cal H}(\eta_1)\Delta\eta_j +
\left({\cal H}^2(\eta_1)-{1\over 2}{a''(\eta_1)\over a(\eta_1)}\right)\Delta\eta_j^2
-\psi(\eta_j,\bm x(\eta_j))
+\bm \gamma\cdot {\bm V}+(\bm \gamma\cdot {\bm V})^2+{1\over 2}|\bm V|^2
\biggr\},
\end{eqnarray}
where $\bm V$ is the peculiar velocity, which should be understood as 
$\bm V=\bm V(\eta_j,\bm x(\eta_j))$.
In the case of a void, we include the Hubble term because the void is a cosmological
structure distributed on larger scales compared with a cluster of galaxies.

By assuming spherical symmetry of the system statistically, the linear terms
in $\bm V$ and $\Delta \eta_j$ vanish, i.e., $\int d\chi n_{\rm g} \bm \gamma\cdot \bm V=\int d\chi n_{\rm g} \Delta \eta_j=0$, and we have
\begin{eqnarray}
1+\langle z(x_\perp) \rangle &=& 1+z_1+(1+z_1){{ \int d\chi n_{\rm g}(\chi,x_\perp)
\widetilde\delta z(\chi,x_\perp) }
\over {\int d\chi n_{\rm g}(\chi,x_\perp)}}
\label{deltazvoid}
\end{eqnarray}
with 
\begin{eqnarray}
\widetilde\delta z(\chi,x_\perp) &=&
\biggl\{\left({\cal H}^2(\eta_1)-{1\over 2}{a''(\eta_1)\over a(\eta_1)}\right)\Delta\eta_j^2
-\psi(\eta_j,\bm x(\eta_j))
+(\bm \gamma\cdot {\bm V})^2+{1\over 2}|\bm V|^2
\biggr\},
\end{eqnarray}
where we should understand that $\bm x=(\chi,\bm x_\perp)$.
We need to perform the projection along the line-of-sight direction, i.e., integration of
$\widetilde\delta z(\chi,x_\perp)$ over the line-of-sight coordinate $\chi$
in some range with fixed $x_\perp$, and we consider the relative gravitational redshift
defined by
\begin{eqnarray}
  \langle z(x_\perp)\rangle -\langle z(0)\rangle
 = (1+z_1){{ \int d\chi n_{\rm g}(\chi,x_\perp)
\widetilde\delta z(\chi,x_\perp) }
\over {\int d\chi n_{\rm g}(\chi,x_\perp)}}
-(1+z_1){{ \int d\chi n_{\rm g}(\chi,0)
\widetilde\delta z(\chi,0) }
\over {\int d\chi n_{\rm g}(\chi,0)}}.
\label{deltazvoid}
\end{eqnarray}

\begin{figure}[t]
\begin{center}
    \includegraphics[width=8cm]{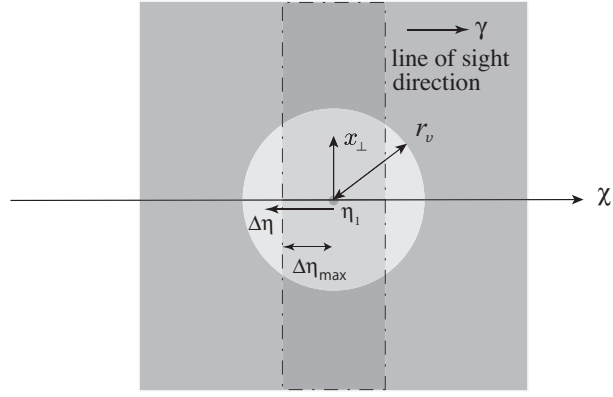}
\caption{Galaxies in the region surrounded by the dash-dotted line 
  are used for the (projection) analysis, and we perform the integration over $\chi$
  (projection along the line-of-sight direction) 
  by fixing $x_\perp$, where $\chi$ and $\Delta \eta$ are related by 
$\chi=\eta_0-\eta=\eta_0-\eta_1-\Delta\eta$.
\label{fig:configurationsint}}
\end{center}
\end{figure}

\red{In the latter part of this section, we demonstrate a possible signal of the
gravitational potential in the redshift of galaxies associated with voids.}
We here adopt the simple model for a spherically symmetric void in Ref.~\cite{VIMOSP}, 
where the integrated density contrast of matter is given in the form
\begin{eqnarray}
\Delta(r)=\Delta_c e^{-(r/r_v)^\alpha},
\end{eqnarray}
where $\Delta_c$, $r_v$, and $\alpha$ are the parameters.
$\Delta_c$ specifies the amplitude of the density contrast, $r_v$ is the characteristic
radius, and $\alpha$ characterises the steepness of the void wall.
\red{This void profile is quite simple, however, it is used in the
  analysis of voids in Ref.~\cite{VIMOSP}, which demonstrates that it works in
  a practical analysis.} 
In the present paper, we adopt $\Delta_c=-0.8$ and $\alpha=3$.
Here $\Delta(r)$ is related to the matter density contrast $\delta(r)$ 
and the gravitational potential $\psi$ by 
\begin{eqnarray}
&&\Delta(r)={3\over r^3}\int_0^r dr'r'{}^2 \delta(r),
\\
&&\triangle \psi(r)=4\pi G a^2\bar\rho_{\rm m}(a)\delta(r),
\end{eqnarray}
where $\bar \rho_{\rm m}(a)$ is the background matter density. 
Assuming a spatially flat cosmology with a cosmological constant, 
we may write $\bar \rho_{\rm m}(a)=8\pi G\Omega_mH_0^2/3a^3$, where 
$\Omega_m$ is the density parameter and $H_0$ is the Hubble parameter 
at the present epoch. Then, the density contrast and the gravitational 
potential of the model are given by
\begin{eqnarray}
&&\delta(r)={1\over r^2}{d\over dr}\left({r^3\Delta(r)\over 3}\right)=
\Delta_c \left(1-{\alpha\over 3}\left({r\over r_v}\right)^\alpha\right)e^{-(r/r_v)^\alpha},
\\
&&\psi(r) =-{3\Omega_m\over 2a}H_0^2
\int_r^\infty dr'r'{ \Delta(r')\over 3} 
=-{H_0^2r_v^2\over 2}{\Omega_m\Delta_c\over \alpha a}\Gamma(2/\alpha,(r/r_v)^\alpha),
\end{eqnarray}
where $\Gamma(z,a)$ is the incomplete Gamma function. 

By solving the continuity equation, the peculiar velocity of the radial direction 
can be written (see, e.g., Ref.\cite{VIMOSP}) as 
\begin{eqnarray}
&&V(r)=-{\cal H}r
\Delta(r) {f(a)\over3},
\label{PV}
\end{eqnarray}
where $f(a) ={d\ln D_1(a)/d\ln a}$ is the growth rate defined by 
logarithmic differentiation with respect to the scale factor $a$, 
which is approximately written as 
$f(a)=[\Omega_m(a)]^\gamma$ with $\Omega_m(a)={a^{-3}\Omega_m /({a^{-3}\Omega_m}+1-\Omega_m)}$ 
and $\gamma=0.55$. In the present paper, we assume that galaxies follow the matter
peculiar velocity field. 
\begin{figure}[t]
\begin{center}
    \includegraphics[width=8.5cm]{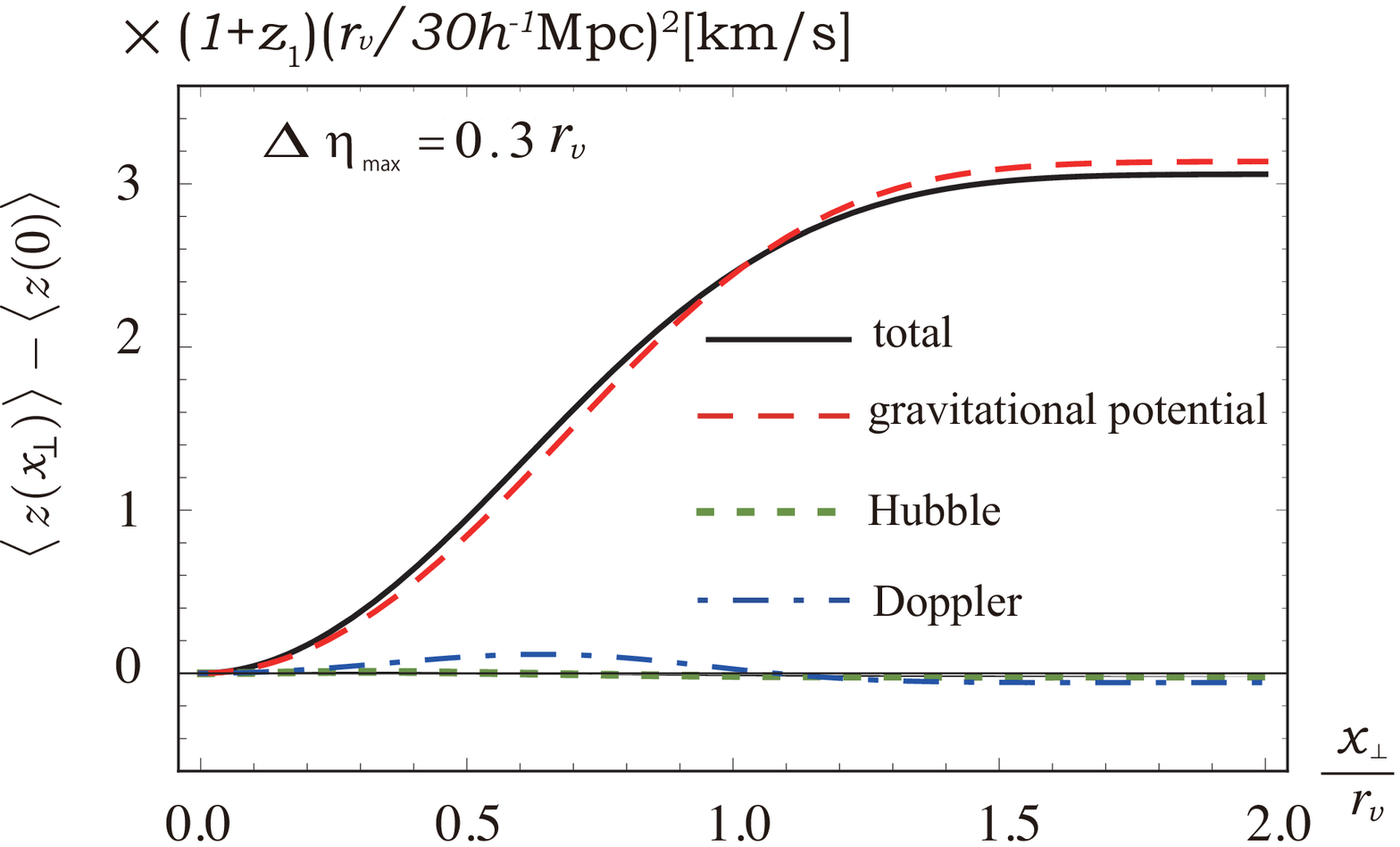}
~~~~~
    \includegraphics[width=8.5cm]{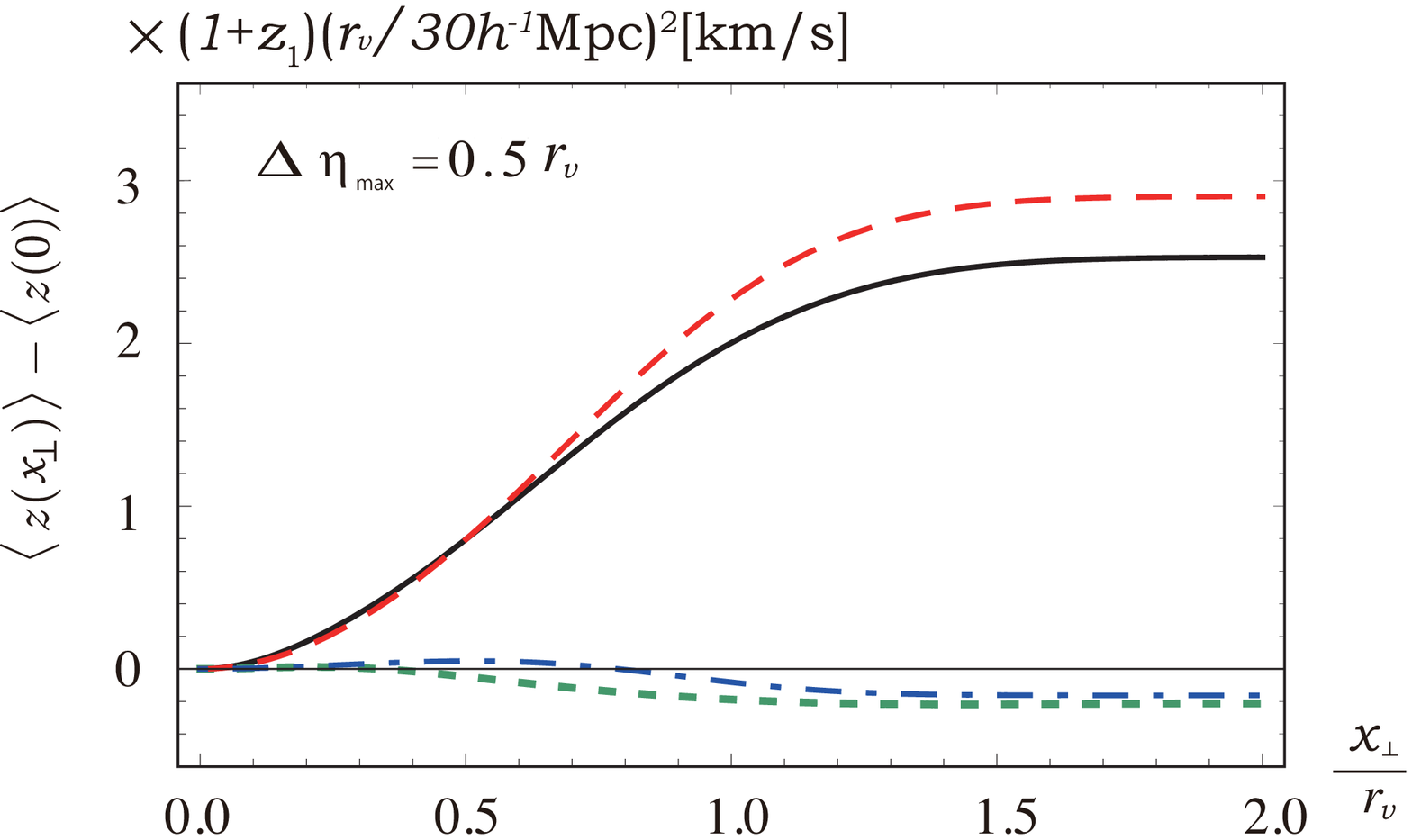}
\\
\vspace{7mm}
    \includegraphics[width=8.5cm]{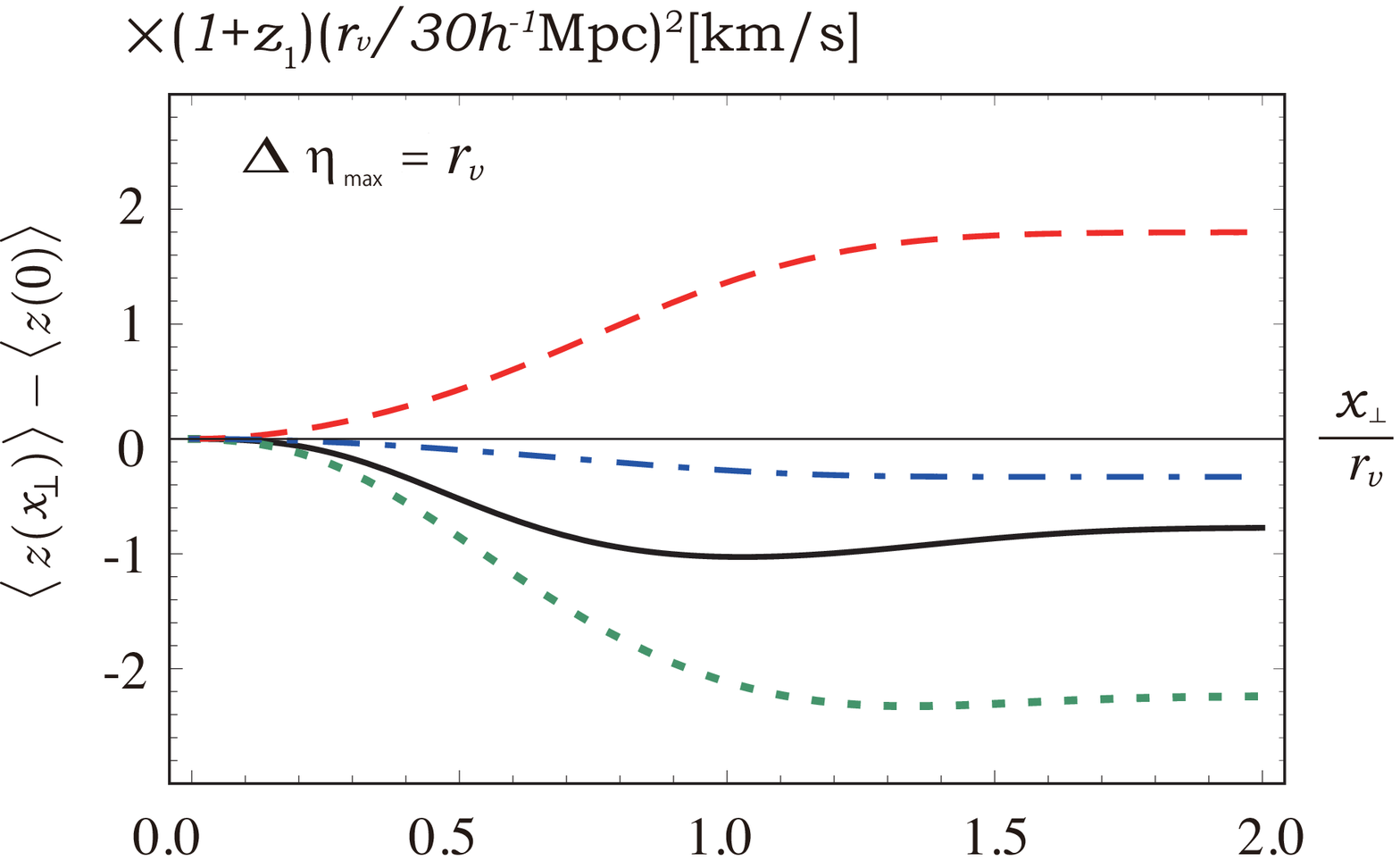}
~~~~~
    \includegraphics[width=8.5cm]{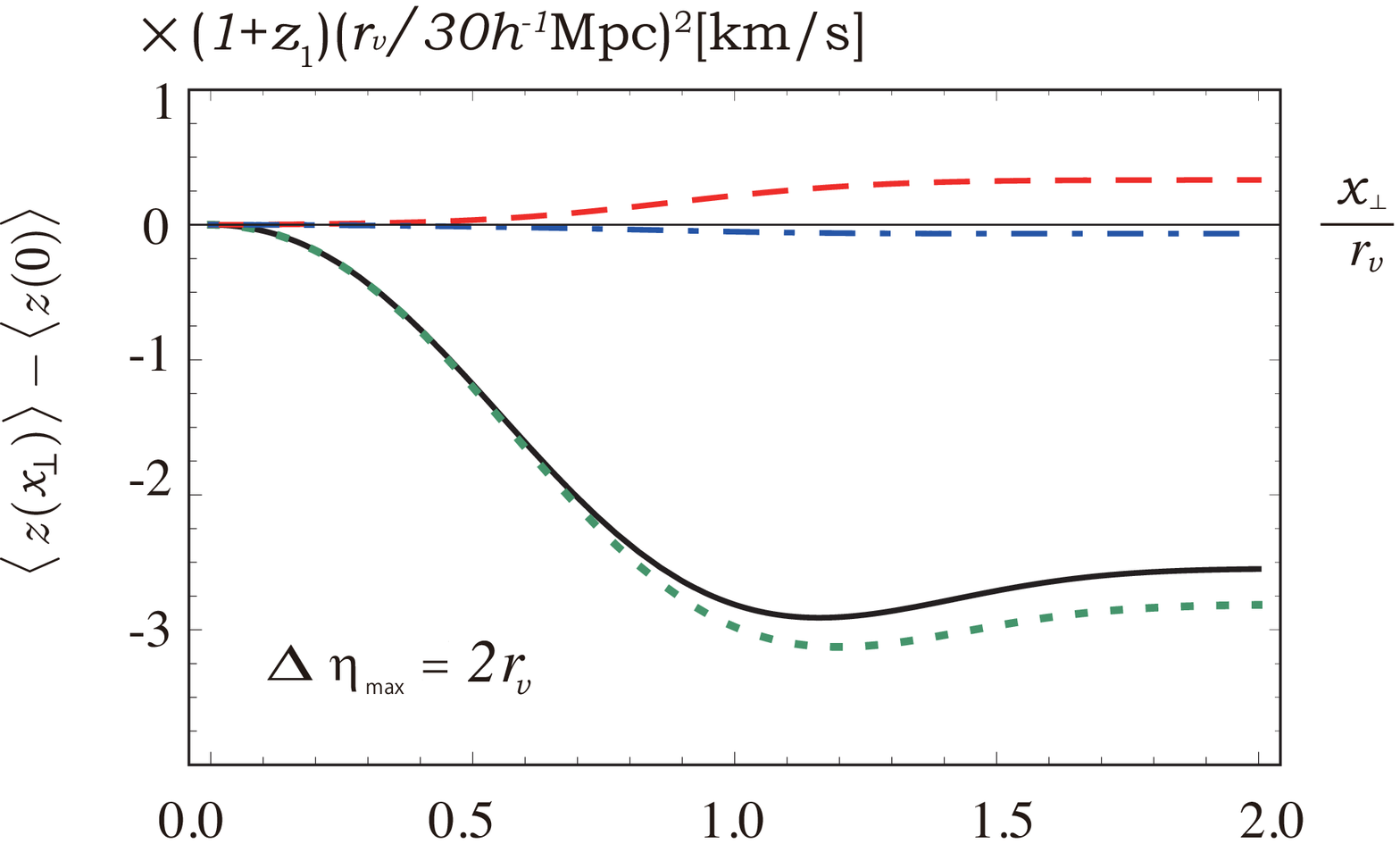}
\caption{Relative gravitational redshift 
$\langle z(x_\perp)\rangle-\langle z(0)\rangle$ in the unit of km/s,
which is  normalised by $(1+z_1)(r_v/30h^{-1}{\rm km/s})^2$ 
  as functions of  $x_\perp /r_v$. 
The black solid curve is the total of $\langle z(x_\perp)\rangle-\langle z(0)\rangle$, 
the green dotted curve is the second-order Hubble term,  
the red dashed curve is the gravitational potential term, 
and the blue dash-dotted curve is the Doppler term contribution.  
In these plots, we adopted different ranges of the projection
along the line-of-sight coordinate $\eta_1-\Delta \eta_{\rm max}<\eta<\eta_1+\Delta \eta_{\rm max}$
(see Fig.~\ref{fig:configurationsint}) with $\Delta \eta_{\rm max}=0.3 r_v$ (upper left panel), 
$\Delta \eta_{\rm max}=0.5 r_v$ (upper right panel), 
$\Delta \eta_{\rm max}=r_v$ (lower left panel), and 
$\Delta \eta_{\rm max}=2r_v$ (lower right panel),  respectively, 
where  we adopted the parameters $\alpha=3$, $\Delta_c=-0.8$, $\Omega_m=0.3$, and $z_1=1$.
\label{fig:results}}
\end{center}
\end{figure}

Figure \ref{fig:results} shows each term in the expression of
the relative gravitational redshift of Eq.~(\ref{deltazvoid}) in the unit of km/s,
which is normalised by $(1+z_1)(r_v/30h^{-1}{\rm Mpc})^2$ as functions of $x_\perp /r_v$. 
In each panel, the red dashed curve is the gravitational potential term,
the green dotted curve is the Hubble term, blue dash-dotted is the Doppler term,
and the black solid curve is the total gravitational redshift.
We find that the amplitude of the gravitational redshift of voids is 
\begin{eqnarray}
\langle z(x_\perp)\rangle-\langle z(0)\rangle
\sim -{\cal O}(0.1)\times (H_0r_v)^2(1+z_1)\sim -{\cal O}(10^{-6})
\times\left({r_v\over 10\;h^{-1}{\rm Mpc}}\right)^2(1+z_1).
\label{gred2}
\end{eqnarray}
The signal changes depending on the range of the projection.
Each panel of Fig.~\ref{fig:results} depicts a different range
of the projection along the line-of-sight direction.
When the range of the projection is narrow,
the gravitational potential term dominates the gravitational redshift. 
However, when the range of the projection is wide, the second-order Hubble term
makes a large contribution. 
Thus, the amplitude of the signal of the gravitational redshift of voids
changes depending on the range of the projection along the line-of-sight direction.
\red{When the background expansion of the universe is well determined, we might be able
  to subtract the dominant contribution from the second-order Hubble term in an analysis
  with some calibration technique.}

\red{From Fig.~\ref{fig:results}, we find that the contribution from second order Doppler term
(blue dash-dotted curve) is quite smaller than that from the gravitational potential
  term (red dashed curve). The amplitude of the gravitational potential term is typically
  several times larger than that of the second order Doppler term.
  Furthermore, the amplitude of the velocity of void is zero at the centre of voids,
  while the gravitational potential $\psi$ has a finite value at the centre of voids.
  The integration of each term over $\chi$ makes the
large difference between these two contributions in Fig.~\ref{fig:results}.}

\red{The gravitation redshifts of the order of a few km/s have been measured in 
the previous works using $10^5$ galaxies associated with clusters \cite{Wojtak,Zhao,Jimeno}.
Then, we think that it might be possible to detect the gravitational redshift of voids.
However, the absence of a galaxy at the center of a void might cause a difficulty
in the detection of the gravitational redshift of voids. In the case of cluster of
galaxies, a central galaxy can be used to measure the relative redshift.
In contrast, we cannot use such an object at the center of void.
This is a difference which might make difficult to detect the gravitational
redshift of voids. In the analysis in this section, we considered a simple
method of averaging the redshifts of galaxies. 
A more sophisticated method to detect the signal of the gravitational
potential of voids might be developed, although such investigation
is beyond the scope of the present paper. }

\section{Summary and Conclusions}
We have investigated possible signals for the gravitational redshift 
in clusters and voids. 
Galaxies associated with clusters are the most promising objects for detecting the
gravitational redshift, as demonstrated in previous works. With the use of
the HOD description with central galaxies and satellite
galaxies in a redshift survey, we have investigated the gravitational redshift 
of satellite galaxies virialised in halos relative to those of the central galaxies. 
In this model, the satellite galaxies are restricted to those located within the virial
radius, which limits the information available compared with that of previous works. 
Our simple analytic model is useful for understanding how the gravitational
redshift signal depends on the HOD properties of galaxy samples.
The virialised random motions of satellite galaxies in halos makes a large
contribution to the gravitational redshift through the second-order Doppler effect.
This feature is potentially useful for testing modified gravity models. 

We have also investigated the gravitational redshift in measurements of intracluster gas. 
Developing a simple model for the intracluster gas including the nonthermal pressure
generated from the random motions proposed by numerical simulations, 
we evaluated a possible signal of the gravitational redshift of intracluster gas. 
The gravitational redshift is dominated by the gravitational potential term, but 
the nonthermal pressure term makes a slight contribution. 
For a detection of the relative gravitational redshift, measurements of the outskirt
region are essential.

Finally, we have investigated the gravitational redshift of voids.
Adopting a very simple model of a void profile, we obtained an analytic formula
for the gravitational redshift.
The amplitude of the signal is \red{$\delta z={\cal O}(1)$--$
{\cal O}(10)$ km/s} depending on the size of the void.
The signal of the relative gravitational redshift depends on the range of the
projection of galaxies along the line-of-sight direction. 
When the range of the projection is narrow, the gravitational potential term
dominates the gravitational redshift. However, when the range of the projection
is wide, the second-order Hubble term makes a large contribution.
These results should be tested more carefully using mock catalogs and galaxy samples,
including estimations of statistical and systematic errors.

\vspace{2mm}
\section*{Acknowledgments}
\red{
We thank anonymous referee for the crucial comments on our first version of the manuscript,
including pointing out our misunderstanding, which significantly improved this paper. }
This work is supported by MEXT/JSPS KAKENHI Grant Numbers 15H05895, 
17K05444, and 17H06359 (KY). We thank N. Okabe, N. Werner, and Y. Fukazawa,
B. Granett for useful communications.
We also thank A. Taruya, S. Saito, N. Sugiyama, K. Koyama, 
D. Parkinson, and M. Sasaki for useful discussions and comments in the workshop
YITP-T-17-03. 

\end{widetext}

\appendix

\section{Model of intracluster gas}
\label{sec:gas}
We assume the following equation of state for the thermal gas components in a cluster:
\begin{eqnarray}
 P_{\rm thermal}(r)=n_{\rm gas}(r)T_{\rm gas}(r),
\end{eqnarray}
where we use a $\beta$-model for the three-dimensional electron number density profile \cite{ngas},
\begin{eqnarray}\label{ngas}
 n_{\rm gas}(r)=n_0\left[1+\left(\frac{r}{r_{\rm c}}\right)^2\right]^{-3\beta/2},
\end{eqnarray}
and the three-dimensional temperature profile, 
\begin{eqnarray}
 T_{\rm gas}(r)=
\left\{
\begin{array}{ll}
T_0\left[1+A\left({r}/{r_0}\right)\right]^b & ({\rm Coma~Cluster}),
\\
~  & ~
\\
T_0\frac{
\left({r}/{r_{\rm cool}}\right)^{a_{\rm cool}}+{T_{\rm min}}/{T_0}}
              {1+\left({r}/{r_{\rm cool}}\right)^{a_{\rm cool}}}
         \frac{\left({r}/{r_{\rm t}}\right)^{-a}}{\left[1+\left({r}/{r_{\rm t}}\right)^b\right]^{c/b}} & ({\rm Perseus~Cluster}).
\end{array}
\right.
\label{tgas}
\end{eqnarray}
In the appendix the temperature of gas is denoted by $T_{\rm gas}(r)$ instead of $T(r)$, following the
previous works (e.g., \cite{Telkina1}).
The fitting functions in Eq.~(\ref{tgas}) are from Refs.~\cite{tgas1} and \cite{tgas2} for the
Coma Cluster and the Perseus Cluster, respectively, and the parameters are listed in Table VI
and Table VII.

For the matter distribution, we assume the Navarro--Frenk--White (NFW) profile, Eq.~(\ref{NFWprofile}),
and the integrated mass within the radius $r$ is given by
\begin{eqnarray}
 M(<r)=4\pi\int_0^r drr^2\rho_{\rm NFW}(r)
       =4\pi\rho_{\rm s}r_{\rm s}^3\left[\ln(1+r/r_{\rm s})-\frac{r/r_{\rm s}}{1+r/r_{\rm s}}\right],
\end{eqnarray}
which gives the gravitational potential by solving
\begin{eqnarray}
 {d\psi(r)\over dr}=\frac{GM(<r)}{r^2}.
\end{eqnarray}
As described in Sec.~III, we introduce the concentration parameter $c$ and the 
virial mass $M_{\rm vir}$ instead of $\rho_s$ and $r_s$. Here we follow the 
definition $M_{\rm vir}=M(<r_{\rm vir})={4\pi}r_{\rm vir}^3\Delta_{\rm c}\rho_{\rm c}/3$,
where $\rho_{\rm c}$ is the critical density, and we adopt $\Delta_{\rm c}=100$, 
determined by the spherical collapse model \cite{Nakamura}.

\begin{figure}[t]
\begin{center}
\vspace{0cm}
 \includegraphics[width=150mm]{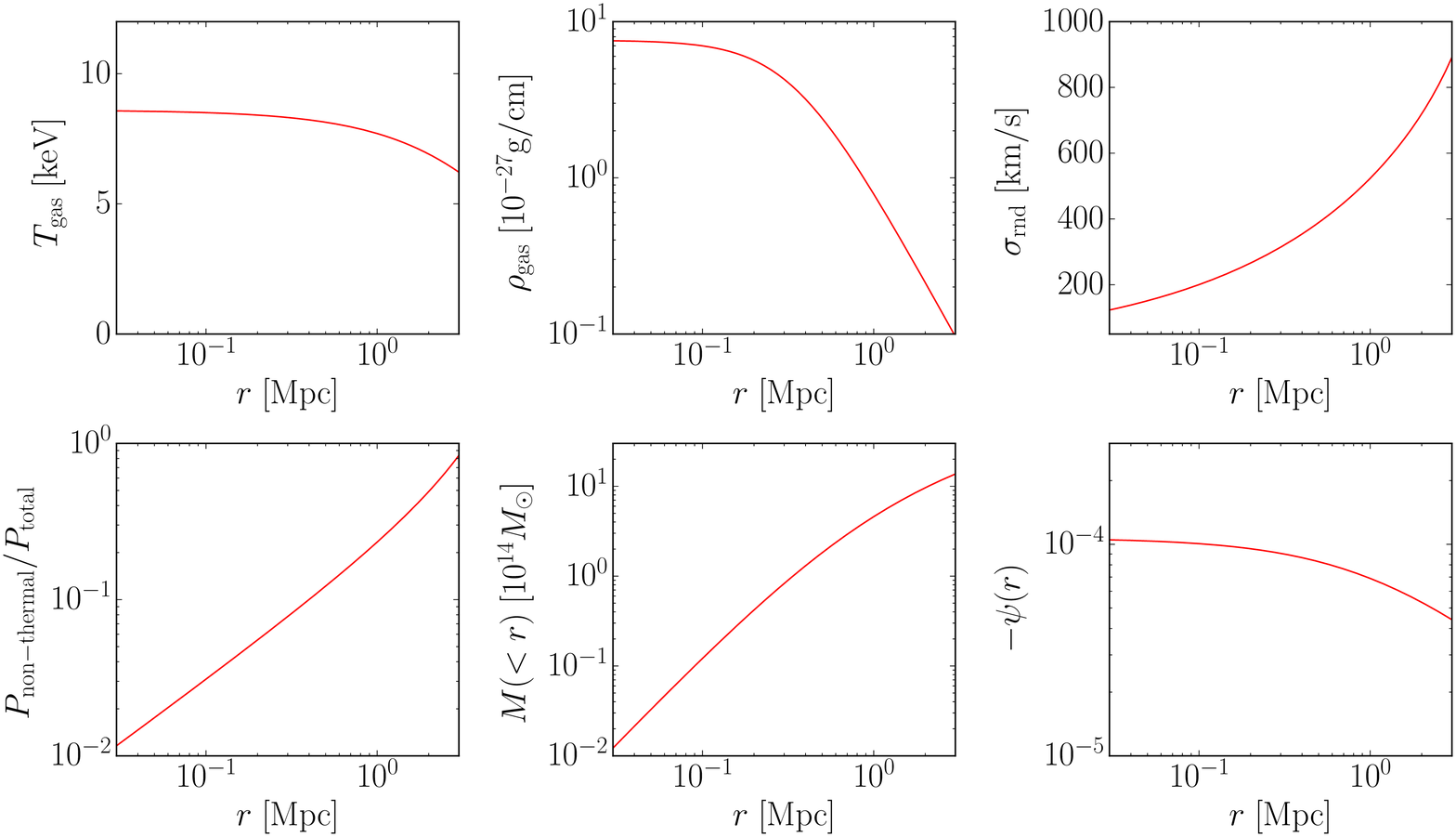}
\vspace{0cm}
\end{center}
 \caption{
   Three-dimensional profiles of the various quantities for the Coma Cluster
   used for Fig. \ref{fig:GR}.
}
\label{fig:3dcoma}
\end{figure}

\begin{figure}[t]
\begin{center}
\vspace{0cm}
 \includegraphics[width=150mm]{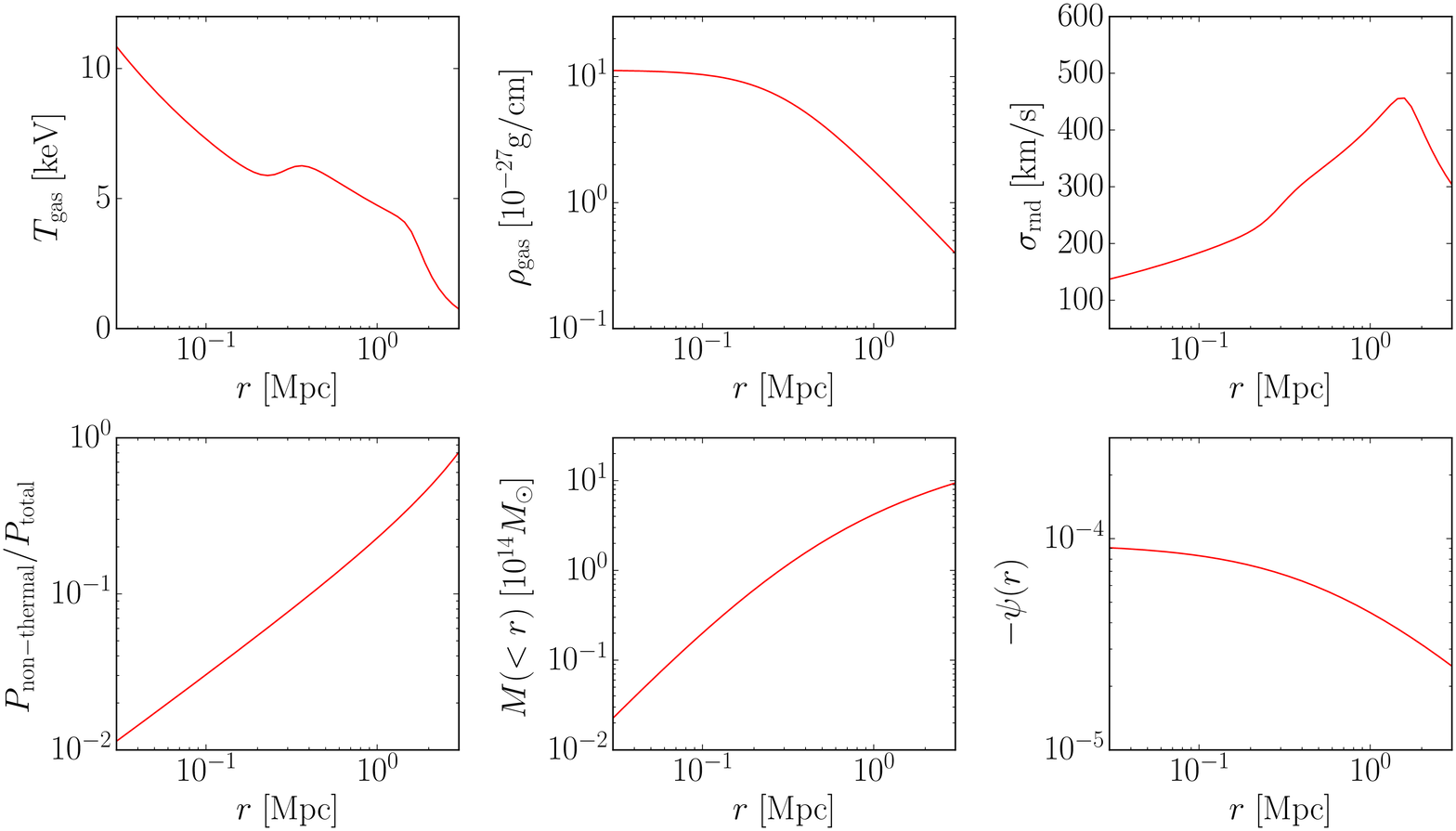}
\vspace{0cm}
\end{center}
 \caption{
   Same as Fig.~\ref{fig:3dcoma} but for the Perseus Cluster used for
   Fig. \ref{fig:GR_per}.
}
\label{fig:3dperseus}
\end{figure}

For the demonstration in Sec.~V, we simply adopt the fitted parameters obtained in previous works for
 the parameters in Eqs.~(\ref{ngas}), (\ref{tgas}),  and (\ref{NFWprofile}).
The numerical values of the parameters are listed in Tables \ref{parameters_coma} 
and \ref{parameters_perseus} for the Coma Cluster and the Perseus Cluster, respectively.
The panels in Figs.~\ref{fig:3dcoma} and \ref{fig:3dperseus} show the three dimensional profile 
of the temperature $T_{\rm gas}(r)$, the gas mass density $\rho_{\rm gas}(r)$, the variance of the 
random velocity $\sigma_{\rm rnd}^2(r)$, the fraction of the nonthermal pressure to the total 
pressure $P_{\rm non-thermal}(r)/P_{\rm total}(r)$, the integrated mass $M(<r)$, and 
the gravitational potential $|\psi(r)|$. 

\begin{table}[!h]
\caption{
 Fitted parameters for the Coma Cluster given in the previous works. 
 Listed are the gas number density profile, Eq.~(\ref{ngas}) (left table),
 the gas temperature profile, Eq.~(\ref{tgas}) (middle table), in Ref.~\cite{Telkina1},
 and the NFW profile, Eq.~(\ref{NFWprofile}) (right table), in Ref.~\cite{Okabe}.
\label{parameters_coma}
}
 \begin{tabular}{lll}
  \renewcommand{\arraystretch}{1.5}
  \begin{tabular}[t]{c|c}
   \hline\hline
   ~~~~$n_{\rm gas}$~~~~ &  ~~~~Terukina et al.~\cite{Telkina1}~~~~\\
   \hline
   $n_0$         & --- \\
   $r_{\rm c}$         & ${0.34} ~{\rm Mpc}$               \\
   $\beta$         & ${0.67}$\\
       \hline\hline
  \end{tabular}
      &
  \renewcommand{\arraystretch}{1.5}
  \begin{tabular}[t]{c|c}
   \hline\hline
   ~~~~$T_{\rm gas}$~~~~ &  ~~~~Terukina et al.~\cite{Telkina1}~~~~\\
   \hline
   $T_0$         & ${8.6} ~{\rm keV}$               \\
   $A$       & ${0.082}$               \\
   $r_{\rm 0}$         & ${3.9} ~{\rm Mpc}$               \\
   $b$         & ${5.3}$                   \\
   \hline\hline
  \end{tabular}
  &
      \renewcommand{\arraystretch}{1.5}
      \begin{tabular}[t]{c|c}
       \hline\hline
       ~~~~$\rho_{\rm NFW}$~~~~ &  ~~~~Okabe et al.~\cite{Okabe}~~~~\\
       \hline
       $M_{\rm vir}$   & ${8.95} ~{\times 10^{14}\;h^{-1}M_\odot}$               \\
       $c$         & ${3.5}$               \\
       \hline\hline
      \end{tabular}
 \end{tabular}
\end{table}

\begin{table}[!h]
\caption{
 Fitted parameter values for the Perseus Cluster given in previous works.
 Listed are the intracluster gas number density profile, Eq.~(\ref{ngas}) (left table),
 the gas temperature profile, Eq.~(\ref{tgas}) (middle table), in Ref.~\cite{Urban},
 and the NFW profile, Eq.~(\ref{NFWprofile}) (right table),  in Ref.~\cite{Simionescu}.
\label{parameters_perseus}
}
 \begin{tabular}{lll}
  \renewcommand{\arraystretch}{1.5}
  \begin{tabular}[t]{c|c}
   \hline\hline
   ~~~~$n_{\rm gas}$~~~~ &  ~~~~Urban et al.~\cite{Urban}~~~~\\
   \hline
   $n_0$         & --- \\
   $r_{\rm c}$         & ${0.285} ~{\rm Mpc}$               \\
   $\beta$         & ${0.71}$\\
       \hline\hline
  \end{tabular}
      &
  \renewcommand{\arraystretch}{1.5}
  \begin{tabular}[t]{c|c}
   \hline\hline
   ~~~~$T_{\rm gas}$~~~~ &  ~~~~Urban et al.~\cite{Urban}~~~~\\
   \hline
   $T_0$         & ${4.06} ~{\rm keV}$               \\
   $T_{\rm min}$       & ${2.92} ~{\rm keV}$               \\
   $r_{\rm cool}$         & ${0.29} ~{\rm Mpc}$               \\
   $a_{\rm cool}$         & ${6.72}$                    \\
   $r_{\rm t}$         & ${1.6} ~{\rm Mpc}$               \\
   $a$         & ${0.33}$                    \\
   $b$         & ${16.24}$                   \\
   $c$         & ${2.36}$                  \\
   \hline\hline
  \end{tabular}
  &
      \renewcommand{\arraystretch}{1.5}
      \begin{tabular}[t]{c|c}
       \hline\hline
       ~~~~$\rho_{\rm NFW}$~~~~ &  ~~~~Simionescu et al.~\cite{Simionescu}~~~~\\
       \hline
       $M_{\rm vir}$   & ${8.05} ~{\times 10^{14}\;M_\odot}$               \\
       $c$         & ${6.6}$               \\
       \hline\hline
      \end{tabular}
 \end{tabular}
 
\end{table}




\begin{thebibliography}{99}
\bibitem{Bartolo2016}
N. Bartolo et al., Phys. Dark Univ. {\bf 13} (2016) 30
\bibitem{Baccanelli2016a}
A. Raccanelli, D. Bertacca, D. Jeong, M. C. Neyrinck, A. S. Szalay,  arXiv:1602.03186
\bibitem{Baccanelli2016b}
A. Raccanelli, F. Montanari, D. Bertacca, O. Dore,  R. Durrer, 
J. Cosmol. Astropart. Phys. 05(2016)009 
\bibitem{Bertacca2017}
D. Bertacca et al., arXiv:1705.09306
\bibitem{Yoo2014}
J. Yoo, Class. Quant. Grav. {\bf 31} (2014) 234001
\bibitem{Yoo2009}
J. Yoo, A. L. Fitzpatrick, M. Zalarriaga, Phys. Rev. D {\bf 80} 083514 (2009) 
\bibitem{Alam1}
S. Alam, et al., Mon. Not. R. Astron. Soc. {\bf 470} 2822 (2017)
\bibitem{Alam2}
H. Zhu, et al., Mon. Not. R. Astron. Soc. {\bf 471} 2345 (2017)
\bibitem{Wojtak}
R. Wojtak, S. H. Hansen, J. Hjorth, Nature {\bf 477} 567 (2011)
\bibitem{Zhao}
H. Zhao, J. A. Peacock, B. Li, Phys. Rev. D {\bf 88} 043013 (2013)
\bibitem{Kaiser}
N. Kaiser, Mon. Not. R. Astron. Soc.  {\bf 435} 1278 (2013)
\bibitem{Jimeno}
P. Jimeno, T. Broadhust, J. Coupon, K Umetsu, R. Lazkov, Mon. Not. R. Astron. Soc. {\bf 448} 199 (2015)
\bibitem{Cai}
  Y.-C. Cai, N. Kaiser, S. Cole, C. Frenk, Mon. Not. R. Astron. Soc. {\bf 468} 1981 (2016)


  
\bibitem{Hitomi}
 Hitomi Collaboration, Nature {\bf 535} 117 (2016), arXiv:1607.04487
\bibitem{Dodelson}
  S. Dodelson, {\it Modern Cosmology} (Academic Press, 2003)
  

\bibitem{White}
M. White, Mon. Not. R. Astron. Soc. {\bf 321} 1 (2001)
\bibitem{Seljak}
U. Seljak, Mon. Not. R. Astron. Soc. {\bf 325} 1359 (2001)
\bibitem{CooraySheth2002} 
A. Cooray, R. Sheth, {Phys. Rep.} {\bf  372} 1 (2002)
\bibitem{Zheng2005}
Z. Zheng et al., Astrophys. J. {\bf  633} 791 (2005)
\bibitem{ReidSpergel}
B. A. Reid, D. N. Spergel, {Astrophys. J.} {\bf 698} 143 (2009)

\bibitem{HikageYamamoto2013}
  C. Hikage, K. Yamamoto,  J. Cosmol. Astropart. Phys. 08(2013)019
\bibitem{HikageYamamoto2015}
    C. Hikage, K. Yamamoto, Mon. Not. R. Astron. Soc. Lett. {\bf 455} L77 (2015)
\bibitem{HMTS}
 C. Hikage, R. Mandelbaum, M. Takada, D. N. Spergel, Mon. Not. R. Astron. Soc. {\bf 435} 2345 (2013)
\bibitem{Kanemaru}
T. Kanemaru, C. Hikage, G. Huetsi, A. Terukina, K. Yamamoto, Phys. Rev. D {\bf 92} 023523 (2015) 
\bibitem{NFW}
 J. F. Navarro, C. S. Frenk, S. D. White, Astrophys. J. {\bf 490} 493 (1997)

\bibitem{Parejko}
J. K. Parejko et al., {Mon. Not. Roy. Astron. Soc.} {\bf 429} 98 (2013)

\bibitem{Manera}
M. Manera et al., {Mon. Not. Roy. Astron. Soc.} {\bf 428} 1036 (2013)

\bibitem{ShethTormen1999}
R. K. Sheth, G. Tormen, Mon. Not. R. Astron. Soc. {\bf 308} 119 (1999)

\bibitem{ST2} R. K. Sheth, G. Tormen, Mon. Not. Roy. Astron. Soc. {\bf 329} 16 (2002)

\bibitem{Parkinson} H. Parkinson, S. Cole J. Helly, Mon. Not. Roy. Astron. Soc. {\bf 383} 557 (2008)

\bibitem{BroadHurst}
  T. Broadhurst, E. Scannapieco, Astrophys. J. {\bf 533} L93 (2000)
\bibitem{YSS}
K. Yamamoto, H. Sato, N. Sugiyama, Phys. Rev. D {\bf 56} 7566 (1997)
\bibitem{Battaglia}
 N. Battaglia et al., Astrophys. J. {\bf 758} 74 (2012)
\bibitem{Shaw}
  L. D. Shaw et al., Astrophys. J. {\bf 725} 1452 (2010)
\bibitem{Telkina1}
A. Terukina, L. Lombriser, K. Yamamoto, D. Bacon, K. Koyama, R. C. Nichol, 
J. Cosmol. Astropart. Phys. 04(2014)013
\bibitem{Telkina2}
A. Terukina, K. Yamamoto, N. Okabe, K. Matsushita, T. Sasaki, 
J. Cosmol. Astropart. Phys. 10(2015)064
\bibitem{Shi}
 X. Shi, E. Komatsu, K. Nelson, D. Nagai, Mon. Not. R. Aastron. Soc. {\bf 448} 1020 (2015)

\bibitem{UVP}
N. Hamaus, P. M. Sutter, B. D. Wandelt, Phys. Rev. Lett. {\bf 112} 251302 (2014) 
\bibitem{Mao}
Q. Mao et al., Astrophys. J. {\bf 835} 160 (2017).
\bibitem{Micheletti}
D. Micheletti et al., Astron. Astrophys. {\bf 570} A106 (2014)
\bibitem{CaiII}
Y.-C. Cai, N. Padilla, B. Li, Mon. Not. R. Astron. Soc. {\bf  451} 1036 (2015)
\bibitem{VIMOSP}
 A. J. Hawken et al.,  Astronomy and Astrophysics, {\bf 607} A54 (2017) 
\bibitem{Nico}
 N. Hamaus et al., Journal of Cosmology and Astroparticle Physics, 07(2017)014
\bibitem{Achitouv1}
 I. Achitouv, arXiv:1707.08121
\bibitem{ngas}
 A. Cavaliere, R. Fusco-Femiano, Astron.  Astrophys. {\bf 70} 677 (1978)
\bibitem{tgas1}
 J. O. Burns, S. W. Skillman, B. W. O'Shea, Astrophys. J. {\bf 721} 1105 (2010)
\bibitem{tgas2}
 A. Vikhlinin, A. Kravtsov, W. Forman et al., Astrophys. J. {\bf 640} 691 (2006)
\bibitem{Urban}
 O. Urban, A. Simionescu, N. Werner et al., arXiv:1307.3592
\bibitem{Nakamura}
 T. T. Nakamura, Y. Suto, Prog. Theor. Phys. {\bf 97} 49 (1997)
\bibitem{Okabe}
 N. Okabe, Y. Okura, T. Futamase, Astrophys. J. {\bf 713} 291 (2010)
\bibitem{Simionescu}
 A. Simionescu et al., Science {\bf 331} 25 (2011)



\end{thebibliography}
\end{document}